\documentclass[a4paper,fleqn,usenatbib]{mnras}

\usepackage{newtxtext,newtxmath}

\usepackage[T1]{fontenc}
\usepackage{ae,aecompl}

\usepackage{graphicx}	
\usepackage{amsmath}	
\usepackage{amssymb}	

\usepackage{bm}
\usepackage{hyperref}
\usepackage{listings}
\usepackage{pdflscape}

\newcommand{\as}{$^{\prime\prime}$}

\newcommand{\htwo}{H$_2$}

\newcommand{\ls}{L$_{\odot}$}
\newcommand{\ms}{M$_{\odot}$}
\newcommand{\kms}{$\,\rm km\,s^{-1}$}

\newcommand{\eqq}{\!=\!}  
 
\newcommand{\tdep}{$\tau_{\rm dep}$}

\newcommand{\jone}{{$J\eqq$ 1$-$0}} 

\newcommand{\jthree}{{$J\eqq$ 3$-$2}}
\newcommand{\jfour}{{$J\eqq$ 4$-$3}}
\newcommand{\jfive}{{$J\eqq$ 5$-$4}}
\newcommand{\jsix}{{$J\eqq$ 6$-$5}}
\newcommand{\jseven}{{$J\eqq$ 7$-$6}}
\newcommand{\jeight}{{$J\eqq$ 8$-$7}}
\newcommand{\jnine}{{$J\eqq$ 9$-$8}}

\newcommand{\jeleven}{{$J\eqq$ 11$-$10}}

\newcommand{\jthirteen}{{$J\eqq$ 13$-$12}}
\newcommand{\jfourteen}{{$J\eqq$ 14$-$13}}
\newcommand{\jthirty}{{$J\eqq$ 30$-$29}}

\newcommand{\lfir}{$L_{\rm FIR}$}
\newcommand{\lprimeco}{$L^\prime_{\rm CO}$}

\newcommand{\mhtwo}{$M_{\rm H_2}$}
\newcommand{\mstellar}{$M_{*}$}
\newcommand{\alphaco}{$\alpha_{\rm CO}$}
\newcommand{\tkin}{$T_{\rm kin}$}
\newcommand{\nhtwo}{$n_{\rm H_2}$}
\newcommand{\Nco}{$N_{\rm CO}$}

\title[Warm and Cold Molecular Gas in 87 Galaxies]{Warm and Cold Molecular Gas Conditions Modeled in 87 Galaxies Observed by the {\it Herschel} SPIRE Fourier Transform Spectrometer}

\author[J. Kamenetzky et al.]{
J. Kamenetzky,$^{1}$\thanks{E-mail: jkamenetzky@westminstercollege.edu}
N. Rangwala,$^{2}$
and J. Glenn$^{3}$
\\
$^{1}$Westminster College, 1840 South 1300 East, Salt Lake City, UT 84105\\
$^{2}$NASA Postdoctoral Fellow, NASA Ames Research Center/ Oak Ridge Associated Universities, Naval Air Station, Moffett Field, Mountain View, CA 94035\\
$^{3}$University of Colorado Boulder, Center for Astrophysics and Space Astronomy, 389-UCB, Boulder, CO, USA}

\date{Accepted XXX. Received YYY; in original form ZZZ}

\pubyear{2017}

\begin{document}
\label{firstpage}
\pagerange{\pageref{firstpage}--\pageref{lastpage}}
\maketitle

\begin{abstract}
We have conducted two-component, non-LTE modeling of the CO lines from \jone\ through \jthirteen\ in 87 galaxies observed by the {\it Herschel} SPIRE Fourier Transform Spectrometer (FTS). 
We find the average pressure of the cold molecular gas, traced especially by CO \jone, is $\sim 10^{5.0 \pm 0.5}$ K cm$^{-3}$. The mid- to high-J lines of CO trace higher-pressure gas at $10^{6.5 \pm 0.6}$ K cm$^{-3}$; this pressure is slightly correlated with \lfir. 
Two components are often necessary to accurately fit the Spectral Line Energy Distributions (SLEDs); a one-component fit often underestimates the flux of CO \jone\ and the mass. 
If low-J lines are not included, mass is underestimated by an order of magnitude.
Even when modeling the low-J lines alone or using an \alphaco\ conversion factor, the mass should be considered to be uncertain to a factor of at least 0.4 dex, and the vast majority of the CO luminosity will be missed (median, 65\%). 
We find a very large spread in our derived values of \alphaco, though they do not have a discernible trend with \lfir; the best fit is a constant 0.7 \ms / (K \kms pc$^2$), 
with a standard deviation of 0.36 dex, and a range of 0.3-1.6 \ms / (K \kms pc$^2$). 
We find average molecular gas depletion times (\tdep) of $10^{8}$ yr that decrease with increasing SFR.
Finally, we note that the \jeleven/\jone\ line flux ratio is diagnostic of the warm component pressure, and discuss the implications of this comprehensive study of SPIRE FTS extragalactic spectra for future study post-{\it Herschel}.
\end{abstract}




\section{Introduction}\label{sec:intro}

The different energy levels of atomic and molecular processes on the quantum scale allow astronomers to probe the different energy regimes of astrophysical processes on galactic scales. As the ISM itself is composed of multiple phases (ionized, atomic, and molecular), so too can each individual component be broken down by regions of higher and lower excitation. Molecular rotational transitions are abundant in the cool molecular gas component of the interstellar medium (ISM). For the molecular ISM, the workhorse molecule of astronomers is carbon monoxide (CO). With a permanent dipole moment, unlike H$_2$, and a ground level transition \jone\ at $\Delta E_{J\ \eqq\ 1-0} = 5.53$ K, CO is an excellent tracer of cold molecular gas. At higher-$J$ transitions, Earth's atmosphere becomes opaque. From the ground, only the first three transitions are typically accessible, up to upper energy level $E_{J\ \eqq\ 3} = 33$ K, though some detections of \jfour\ and \jsix\ can be made. The first observations of \jsix\ from the ground in three starburst galaxies demonstrated the existence of warmer, denser molecular gas than that traced by \jone\ \citep{Harris1991}, but this gas cannot be studied from the ground with multiple lines or in many galaxies. Throughout this work, we use often mid-$J$ to refer to \jfour\ through \jsix, and high-$J$ for \jseven\ and above, unless otherwise specified.

The only way to access the transitions which trace warmer molecular gas is to get above the atmosphere as did the {\it Herschel} Space Observatory. The {\it Herschel} SPIRE Fourier Transform Spectrometer (FTS) was simultaneously sensitive to the \jfour\ ($E_{J\ \eqq\ 4} = 55$ K) up to the \jthirteen\ ($E_{J\ \eqq\ 13} = 503$ K) line of CO for local galaxies. With this access to higher energy CO transitions, {\it Herschel} gave us the tools to more finely discern the lower and higher excitation components of the multi-phase molecular ISM. To more finely measure the excitation of the molecular ISM is to better understand energy exchange between the star formation and the ISM. Many studies of individual galaxies 
\citep{Glenn2015, Hailey-Dunsheath2012, Israel2014, Kamenetzky2012, Meijerink2013, Papadopoulos2014,Panuzzo2010, Pellegrini2013, Rangwala2011,Rigopoulou2013, Rosenberg2014,Schirm2014,Spinoglio2012,Wu2015}
and surveys 
\citep{Daddi2015, Greve2014, Israel2015, Kamenetzky2014, Kamenetzky2016,Lu2014, Liu2015,Mashian2015,Papadopoulos2010, Papadopoulos2012, Papadopoulos2012a, Pereira-Santaella2013,  Pereira-Santaella2014}
have shown that the higher-J lines of CO arise from warmer, denser gas than the cold gas responsible for the emission of e.g., CO \jone. 
Though this warm gas is only a small fraction ($\sim 10\%$) of the total molecular mass, it is responsible for $\sim 90\%$ of the CO luminosity \citep{Kamenetzky2014}. This high CO luminosity cannot be explained by excitation from the ultraviolet (UV) light from young O and B stars in photon-dominated regions (PDRs); mechanical heating is often required. 
Theorists have attempted to use hydrodynamical galaxy simulations to produce galaxy-integrated SLEDs of CO emission \citep{Narayanan2008b, Olsen2016}, but we still lack a comprehensive picture of the mechanisms responsible for this emission. Furthermore, even higher-J lines of CO than addressed here (above \jthirteen) have been detected e.g. with {\it Herschel} PACS, indicating a third, even higher temperature component of molecular gas \citep[NGC1068,][]{Hailey-Dunsheath2012}. The CO SLEDs from \jfourteen\ through \jthirty\ show a large range in SLED shape, even among similar galaxies \citep{Mashian2015}.

How ubiquitous is this highly excited CO emission that was first detected in 1991? We now see that it arises not just from the most rapidly star-forming local galaxies. This paper is the third in a series, which completes our survey of all extragalactic CO observed with the {\it Herschel} SPIRE Fourier Transform Spectrometer (FTS). In \citet{Kamenetzky2014} we demonstrated the two-component modeling of physical conditions, shown again in this work, for a selection of 21 spectra. In \citet{Kamenetzky2016}, we presented the line fluxes and upper limits for all the extragalactic spectra observed by the {\it Herschel} SPIRE FTS. 
We now apply the aforementioned two-component physical condition modeling to all those galaxy line fluxes.

Though the majority of the methodology for the measurement of the line fluxes and the two-component physical condition analysis is presented in previous papers, we briefly summarize them in Sections \ref{sec:obs} and \ref{sec:analysis}. Our results are presented in Section \ref{sec:disc}, including a discussion of the significance of this warm gas and our work's place in the literature in Section \ref{sec:significance}.

\section{Observations}\label{sec:obs}

We use the CO \jone\ through \jthirteen\ line fluxes reported in Tables 2 and 3 of \citet{Kamenetzky2016}; we direct the reader to that paper for more details of the data processing and line fitting. The \jfour\ through \jthirteen\ lines were measured from the {\it Herschel} SPIRE FTS, after referencing each galaxy's entire spectrum to a common 43\farcs{5} beam size, taking the source/beam coupling factor derived from SPIRE photometry maps into account. 
However, most of the galaxies were point-like compared to the FTS beam, which varies from 45\as to 17\as with frequency in the manner shown in Figure 8 of \citet{Swinyard2014}.
43\farcs{5} was chosen because this is the largest beam size of our CO transitions (for CO \jfour).
Table 2 of \citet{Kamenetzky2016} lists 1523 three sigma detections and 1006 upper limits for 227 pointings total, out of a total of all $~ 300$ archived extragalactic spectra observed by the FTS. 

Our fitting routine determined the probability distribution function of the true line flux given the observed line flux because the fits were heavily influenced by the noise pattern in the spectra. In short, the ``ringing" caused by taking the Fourier Transform of the (finite) interferograms can easily be mistaken for line detections by a least-squares fitting method. 

The \jone\ through \jthree\ lines (low-$J$) were collected from the literature and the Arizona Radio Observatory. These lines were also referenced to the same 43\farcs{5} beam size, using the same source/beam size correction. Table 3 of \citet{Kamenetzky2016} lists 750 such lines. Some galaxies had multiple measurements available for the same line; in our modeling, we include all of those, which all contribute individually to the calculation of the likelihood in comparison to the RADEX models (described next). In subsequent figures and calculations in this paper, when e.g., the CO \jone\ line flux is a variable, we use the weighted average of the available measurements. Also in subsequent analysis, we refer to the ``\lfir\ in the beam." For point source galaxies, this is the total \lfir\ (40-120 $\mu$m) of the galaxy, from Hyperleda.\footnote{\url{http://leda.univ-lyon1.fr/}} For extended galaxies, we reduce the total \lfir\ using the same source/beam coupling factors described above.

\section{Analysis}\label{sec:analysis}

We used the nested sampling algorithm MultiNest \citep{Feroz2009} to compare the measured CO spectral line energy distributions (SLEDs) to the ones produced by a custom version of the non-LTE code RADEX \citep{vanderTak2007} written by Phil Maloney. We also used the PyMultiNest Python wrapper \citep{Buchner2014}. Our custom code, which works with the publicly available RADEX\footnote{\url{http://home.strw.leidenuniv.nl/~moldata/radex.html}} and its Python wrapper PyRadex\footnote{\url{https://github.com/keflavich/pyradex}} is available online under the name PyRadexNest\footnote{\url{https://github.com/jrka/pyradexnest}}. We follow the same procedure as described in more detail in \citet{Kamenetzky2014}. 

Briefly, in RADEX, each component of molecular gas is described by a kinetic temperature (\tkin), volume density of the collision partner with CO, molecular hydrogen (\nhtwo), and a column density of CO (\Nco) per unit linewidth. We also include a fourth parameter, the angular area filling factor ($\Phi < 1$), which scales the fluxes linearly. RADEX uses the first three parameters to determine the level populations of the rotational states of CO and the optical depths in each line. 
The non-LTE treatment is necessary because, at warmer temperatures, the levels will be subthermally populated.
Using an escape probability method, RADEX predicts the line intensities as background-subtracted Rayleigh-Jeans equivalent radiation temperatures. 

In PyRadexNest, for a given set of parameters ${\bm p}$, we calculate the negative log likelihood of the predicted RADEX model $I({\bm p})$ given the measurements $\bm x$ and errors $\bm \sigma$ as $\sum_i [0.5 {\rm ln}(2 \pi) + {\rm ln}(\sigma_i) + 0.5 (x_i-I_i(\bm{p}))^2 \sigma_i^{-2}$]. In the case of the one-component model, we have four free parameters (\tkin, \nhtwo, \Nco, and $\Phi$). In the case of the two-component model, we have eight free parameters, four for each component. In the log likelihood, the sum of the two predicted RADEX models is $I(\bm p)$ for comparison to the data. We conducted one-component modeling for all galaxies with at least four lines (176 galaxies), as well as two-component modeling for all galaxies with at least eight lines (87 galaxies). 
Of these 87 pointings, some are actually duplicates of multiple galaxy systems, and are not truly independent: we include three pointings of Arp299, two of NGC1365, and three of NGC2146.

In addition to the free parameters, we calculate secondary parameters such as the pressure ($P$/k = \tkin $\times$ \nhtwo), the mass (M = A $\Phi$ $N_{\rm CO}$ 1.4 m$_{H_2}$ $X_{CO}^{-1}$), the model flux in each line, and the total CO luminosity from the RADEX model (including higher-J lines than observed). Because each of these is also associated with a likelihood, we can determine the distributions of those parameters as well. We do not correct for extinction by dust at high frequencies, which was found to not significantly change the marginalized parameter distributions \citep[section 3.2.4 of ][]{Kamenetzky2014}.

The pressure, as modeled in RADEX, should be considered an {\it effective}, or {\it total} pressure. Our models do not distinguish between excitation due to thermal vs. non-thermal (macroscopic motion) sources. If thermal excitation dominated, our pressure would be equivalent to the thermal pressure (\nhtwo $k_b$ \tkin), and our measured values of \nhtwo\ and \tkin\ would be descriptive of the thermal properties of the gas. Instead, it is likely that pressure due to high non-thermal velocity dispersions, $\sigma_{NT}$, dominates. Some more complex excitation codes, such as \texttt{DESPOTIC}, enhance the collisional excitation rates in the presence of a non-thermal velocity dispersion using a clumping factor, $f_{cl}$ \citep{Krumholz2013}. One can achieve similar excitation by using no clumping, but a higher density of colliding partners, or a higher temperature. Our temperatures and densities represent such higher, enhanced quantities, to account for the non-uniformity of the gas caused by the non-thermal velocity dispersion. 

Because our temperatures and densities are highly degenerate, and only describe the {\it effective} excitation of the gas (not the true thermal properties), we do not present them here. We focus on the total effective pressure, which we henceforth refer to simply as ``pressure."

We utilize a few assumptions, the same as in K14. The relevant parameter for the calculation of the optical depths and level populations is the column density (\Nco) {\it per unit linewidth}; the resultant fluxes simply scale with linewidth, $\Delta V$. So for comparison to RADEX output, we divide all velocity-integrated line fluxes by the linewidth. For each galaxy, we use the median linewidth reported from ground-based data in Table 3 of \citet{Kamenetzky2016}. If none are available, we choose 200 \kms. Our reported masses and luminosities scale linearly with the chosen linewidth.We adopt a relative abundance of CO to \htwo\ ($X_{\rm CO}$) of $10^{-4}$, and a factor of 1.4 to account for helium and other heavy elements in addition to \htwo. 

We also place some restrictions (priors) on the parameter space. First, to define the two components, we require the first one to be of a cooler temperature and more massive than the second, hence referring to them as the ``cool/cold" and ``warm" components. We set a maximum allowable length, $L_{\rm max} = N_{\rm CO} (n_{\rm H_2} X_{\rm CO} \sqrt{\Phi} )^{-1}$, and mass (equation given above) of the system. The mass limit restricts high combinations of $N_{\rm CO}$ and $\Phi$. The length limit restricts combinations of high $N_{\rm CO}$, low $n_{\rm H_2}$, and low $\Phi$. The length limits were calculated by fitting a two dimensional Gaussian to the SPIRE PSW map. For the longest side of the Gaussian, $\sigma$, we assume the true source size $s$ is $\sqrt{\sigma^2 - 19.3^2}$, because the map is the convolution of the true source and the 19\farcs{3} beam of the photometer. The dynamical mass limit is the mass contained at that length at the linewidth, $M_{\rm max} = \Delta V^2 L_{\rm max} / G$. Any grid point which violates these priors is not included in the likelihood. 
To be used in the likelihood calculation, a given line must have an optical depth from RADEX between -0.9 and 100 \citep{vanderTak2007}.

\citet{Tunnard2016} investigated the recoverability of RADEX input parameters using similar methods (grid and Monte Carlo Markov Chain, MCMC) by creating sample SLEDs in RADEX and then fitting them to try to recover the same parameters which produced the SLEDs. For single-species CO models with 10\% errors on the ``data," they found an uncertainty of about 0.5 dex in the recovered parameters (\tkin, \nhtwo). They attribute this to the degeneracies between the physical conditions and the line ratios, and the slow variation of line ratios across parameter space. These degeneracies, and other uncertainties in our measurements, are reflected in the marginalized parameter distributions and error bars that we show in this work. 
Our median one-sigma values (uncertainty in marginalized parameter) for temperature and density are 0.5 (0.2) dex and 1.2 (0.5) dex for the cold (warm) components. In general, the warm component is fitted by more lines and its slope varies more rapidly with parameter space. The median uncertainties for the pressure are 1.0 and 0.3 dex for the cool and warm components (bottom left panel of Figure \ref{fig:results_hist}). As one can see from the rest of the bottom row in this figure, the total mass (mostly cold), total luminosity (mostly warm), and warm component pressure are generally better constrained by the modeling (considering the assumptions listed above) than the physical conditions of temperature and density (cited in text above, not shown in figure).
The long tail in the cold component luminosity uncertainty (lower right of Figure \ref{fig:results_hist}) is because the majority of the luminosity is in the warm component and fixed by the mid- to high-J CO fluxes.


\section{Results and Discussion}\label{sec:disc}

\begin{figure*}
\includegraphics[width=\textwidth]{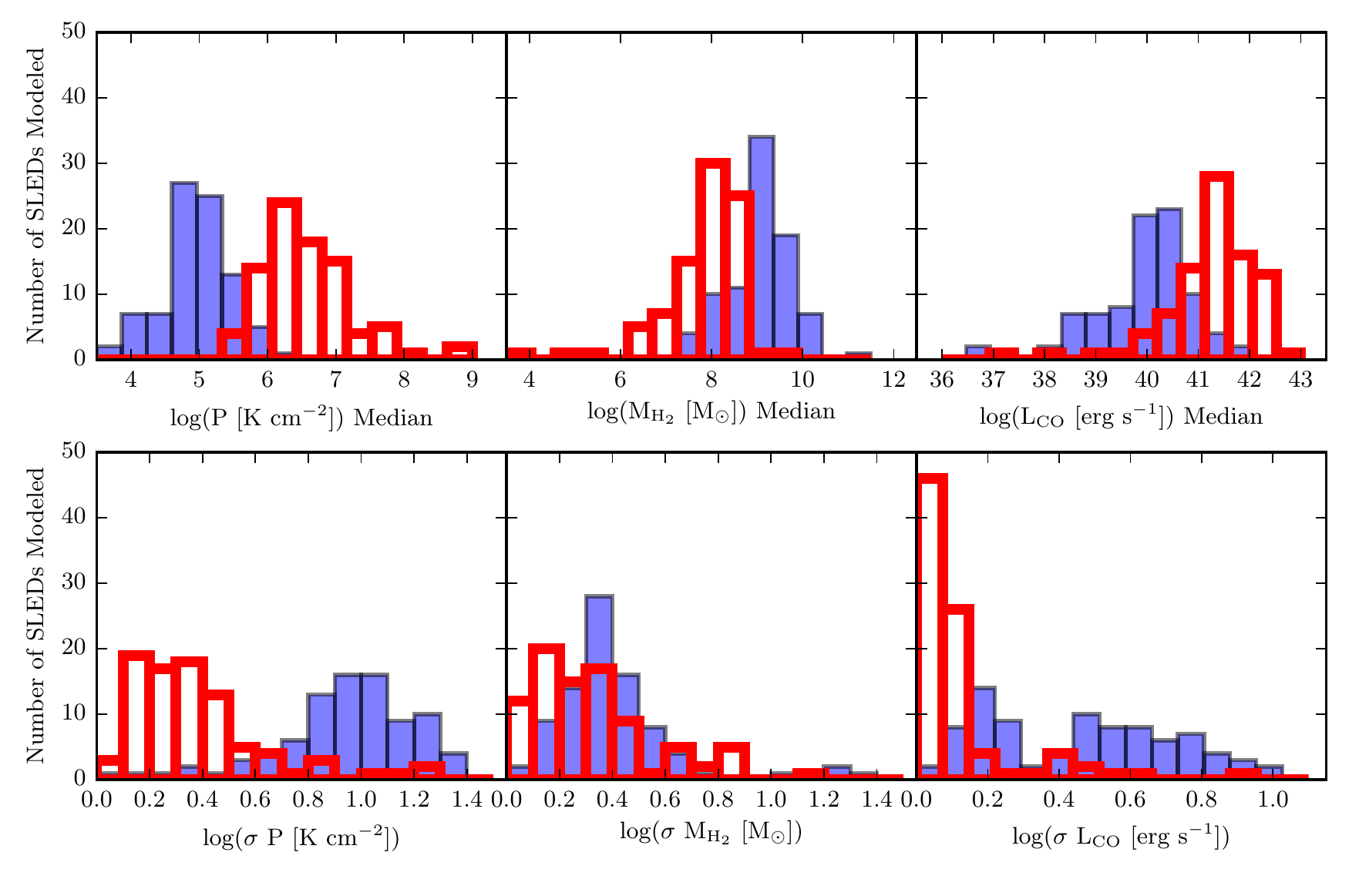} 
\caption{Histogram of Two-Component Modeling Results. The columns are the pressure, mass, and luminosity of the cold (blue, filled) and warm (red, thicker and unfilled) components. The top row show the values of the mode median. The bottom row show the uncertainties (in dex) of the parameters.
\label{fig:results_hist}}
\end{figure*}

In \citet{Kamenetzky2016}, we showed that galaxies with higher \lfir\ (and thus star formation rate, SFR) have higher ratios of high-J to low-J luminosity. For all galaxies (over a couple orders of magnitude in \lfir), however, the ratio of line luminosity to \jone\ luminosity remains relatively flat from \jsix\ to \jthirteen\ (see Figure 2 of the aforementioned paper).
This seems to imply the widespread existence of high pressure molecular gas, but that higher star formation rates may correspond to slightly higher molecular ISM pressures. 
Or, higher star formation rates may indicate a greater fraction of the gas in the high-pressure state.

Though we primarily focus on the results for two-component modeling, it is instructive to compare to three other scenarios: 
\begin{enumerate}
\item modeling all of the same lines using one component, illustrating the necessity of two-component models for accurately describing the SLEDs (``1 Comp"),
\item modeling only the mid- and high-J lines (\jfive\ and above) using one component, as one might do with a high-redshift submillimeter galaxy (``1 Comp, High-J Only"),
\item and modeling only the low-J lines (\jfour\ and below) using one component, as one might have done before {\it Herschel}, with only ground-based data (``1 Comp, Low-J Only") 
\end{enumerate}

We summarize these different scenarios in Table \ref{tbl:summary}. The results of the last two are not shown for individual galaxies. Though the global ISM in a given galaxy is the sum of gas over a gradient of conditions, two discrete components is often the minimum number required to describe the data. See \citet{Kamenetzky2014} for how three (or more) components would overfit the observed  SLEDs.

\begin{table*}
\caption{Summary of Models Being Compared}
\label{tbl:summary}
\begin{tabular}{lllrrrrrr}

\hline
          &                &              & Min. \# & Max. \# & Total \#     &          &                & Also 2 Comp, \\
Models & Lowest & Highest & of lines & of lines & of galaxies & With \jone & Also 2 Comp & with \jone \\
\hline
\hline
2 Comp & \jone & \jthirteen & 8 & 13 & 87 & 78 & -- & -- \\
1 Comp & \jone & \jthirteen & 4 & 13 & 168 & 134 & 87 & 78 \\
1 Comp, High-J Only & \jfive & \jthirteen & 4 & 9 & 128 & 99 & 87 & 78 \\
1 Comp, Low-J Only & \jone & \jfour & 3 & 4 & 99 & 97 & 64 & 63 \\
\hline
\end{tabular}
\end{table*}

The modeling results for individual galaxies are shown in Table \ref{tbl:results} and Figures \ref{fig:A} through A18. 
The results focus not on the primary parameters of the modeling (temperature, density, column density, and filling factor), because those are highly degenerate. Instead, we focus on the pressure (the product of temperature and density), the total mass (proportional to the product of column density and filling factor), and the CO luminosity. 
Figure \ref{fig:results_hist} shows the histograms of these parameters and their uncertainties; in general, the cold component has a median pressure (across the whole sample) of $10^{5.0}$ K cm$^{-2}$, whereas the warm component median pressure is $10^{6.5}$ K cm$^{-2}$. The uncertainty in the cold component pressure is larger, averaging about 1.0 dex instead of 0.3 for the warm component.The histograms of mass and luminosity show the distributions within our sample, and the general comparison between the warm and cold masses (cold is larger) and luminosities (warm is larger). The uncertainty in the warm component luminosity is also much smaller (bottom right panel, ~ 0.1 dex for warm, as opposed to 0.5 dex for cold), as it is well constrained by the slope of the mid- to high-J lines. 
There are simply more lines constraining the warm component fit than the cold component luminosity; see the left column of Figures \ref{fig:A} through A18.
 Perhaps unsurprisingly, the total molecular mass and the CO luminosity are both well-correlated with \lfir, a proxy for SFR. We discuss the correlation between molecular mass and CO \jone\ luminosity in Section \ref{sec:alphaco}.

\subsection{One vs. two-component modeling}\label{sec:components}

\begin{figure}
\includegraphics[width=\columnwidth]{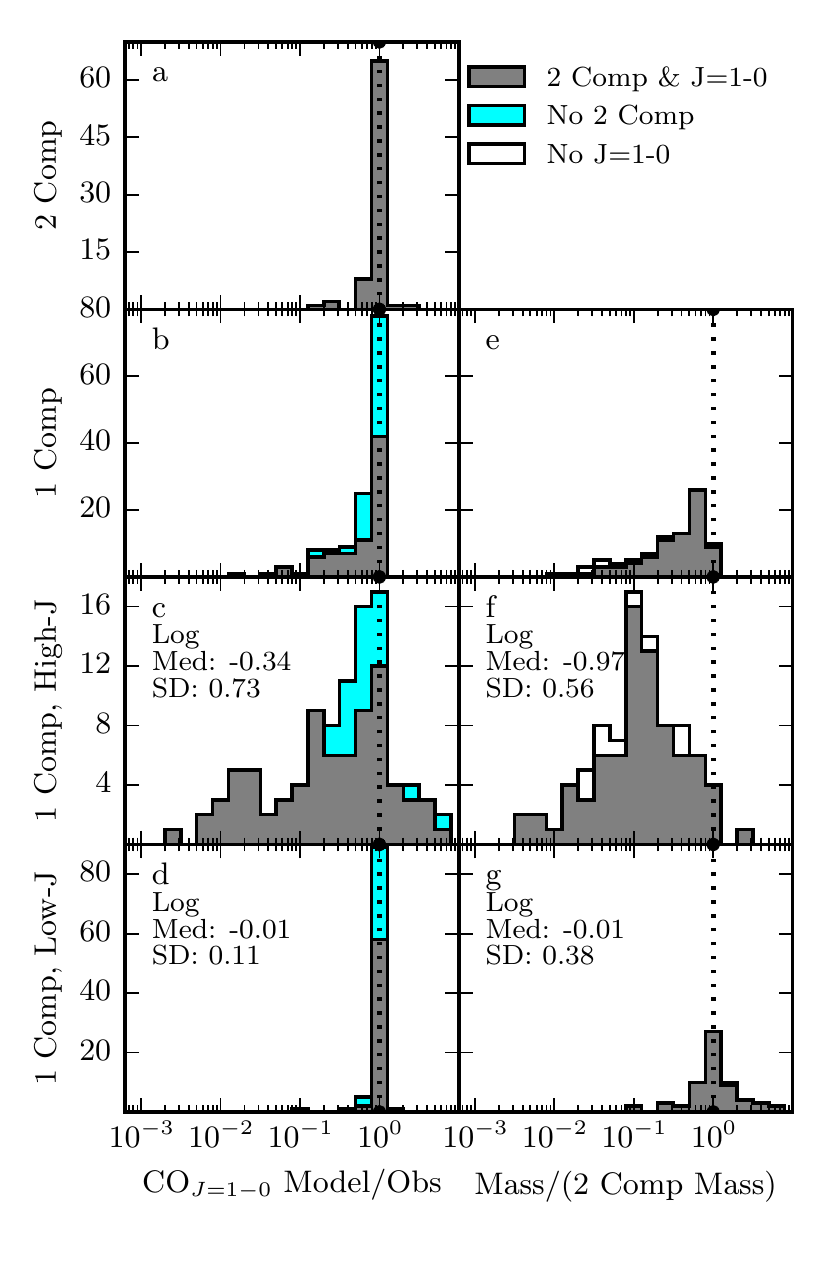} 
\caption{Left Column (panels a-d): Underestimation of the CO \jone\ line flux (and thus mass) in each modeling scenario (see Table \ref{tbl:summary}). The x-axis of these histograms is the ratio of the median modeled CO \jone\ flux from the likelihood to the weighted average of the observed CO \jone\ flux (if at least one measurement is available). 
In the top panel, for two-component modeling, the modeled flux is the sum of the warm and cool component fluxes. 
A ratio of one indicates the model is properly fitting the \jone\ line. 
Right Column (panels e-g): Ratio of modeled 1-component molecular mass and the 2-component modeled mass. The high-J only case (panel f) underestimates the mass by about one order of magnitude, illustrating the importance of including low-J lines for mass estimates and the extent to which high-J lines are tracing a small fraction of the total molecular gas.
Notes for all panels: Galaxies that appear in both the left and right side (which have both a \jone\ measurement for comparison and were modeled with 2 components) are in gray; there are 78 galaxies, except in the bottom panel, where there are 63. The vertical dotted line highlights the 1:1 ratio.
The blue and white are stacked on top of the gray, not behind it.
In the left column, galaxies which were {\it only} modeled as one component (no 2-component modeling was done due to too few lines) are stacked in blue; these galaxies do not appear on the right. Similarly, in the right column, galaxies which do not have a \jone\ line for comparison are stacked in white; these galaxies do not appear on the left.\label{fig:comass}}
\end{figure}

For many of our galaxies, one-component of gas cannot fit the measured CO \jone\ to \jthirteen\ SLED. In the majority of cases where the fit is poor, the fit is driven largely by the slope of the mid- to high-J lines (indicative of the presence of high-pressure molecular gas). The low-J lines (especially CO \jone) are not well fit, and, in fact, are significantly under-predicted.
Rows 12-15 of the Appendix figures suffer from this problem, for example.
We illustrate this point in the left column of Figure \ref{fig:comass}. 
With two-component modeling (top panel, a), the CO \jone\ line is fit within 10\% in 59 of 78 galaxies, or 75\%.
\footnote{
Because the bins of Figure \ref{fig:comass} are in log space, the figure does not directly correspond with the statistics stated in text. The large bin in this figure has 65 galaxies, not 59, because it encompasses ratios from 0.794 to 1.259.
}
(9 galaxies were modeled with two components, but did not have a \jone\ line measurement.) 
In only 4\% (3/78) of the galaxies is the CO \jone\ underestimated by a factor of 2 or more. 
For the same 78 galaxies, when modeled with one component (gray in panel b), the distribution has a long tail. Only 45\% fit the CO \jone\ line within 10\%, and in 32\% of the galaxies the CO \jone\ is underestimated by a factor of two or more. 
When considering all galaxies with at least four lines, when modeled as one-component (gray plus cyan in panel b), the picture is not much better: 47\% fit CO \jone\ within 10\%, 23\% underestimate by a factor of 2 or more. There is not a correlation between this ratio (modeled/observed CO \jone\ flux) and \lfir.
One might expect that higher-SFR galaxies are more likely to require two components to fit the SLEDs. If measuring by the ability to fit the \jone\ line, we do not find this to be the case.

Most of the CO molecules in galaxies are populating the lower energy levels. For cold gas with \jone\ in local thermodynamic equilibrium (LTE), the \jone\ intensity is directly proportional to the mass \citep{Maloney1988,Bolatto2013}. Under such conditions, the fractional underestimation of the \jone\ flux would be equal to the fractional underestimation of the mass. However, we find the underestimation of the mass is larger than that of the \jone\ flux, as shown in the right column of Figure \ref{fig:comass}. In the case of the one-component models using all available lines, though 45\% of the galaxies fit the \jone\ line within 10\% (panel b), only 6\% (5/87) find the same mass as in the two-component model within 10\% (panel e). This is largely due to missing the contribution to the mass from the warm gas and the departure of the \jone\ flux from LTE when the entire SLED must be fit by warmer gas.

A much larger problem occurs, of course, when the CO \jone\ line is not available at all; in this case, the mass is much more uncertain. This is often the case with higher redshift submillimeter galaxies, where only a few (usually mid-$J$) CO lines may be preferentially redshifted into our observable frequency ranges. To estimate the amount by which the CO mass may be underestimated in this case, we compare the high-J only one-component fits (\jfive\ and above, so \jone\ not included) to the two-component fits (which estimate mass well) in panel f of Figure \ref{fig:comass}. On average, the high-J only fit will underestimate the mass by about an order of magnitude. This is consistent with the previously shown result that the warm component mass (traced by high-J) is usually about 10\% of the total molecular mass \citep{Kamenetzky2014}. Futhermore, it illustrates the importance of low-J lines (especially \jone) to mass estimations.

Finally, we compare to the low-J only case (bottom row of Figure \ref{fig:comass}, which is analogous to most of the non-LTE modeling of CO lines from ground-based observatories, before the era of {\it Herschel}. Not surprisingly, when only fitting the first four lines, the CO \jone\ line is extremely well fit (panel d). The mass is not systematically underestimated compared to the 2-component case, but there is some spread (standard deviation 0.4 dex) shown in panel g.
This is consistent with average uncertainty in the mass parameter shown in Figure \ref{fig:results_hist} (bottom middle) of about 0.4 dex, or about a factor of 2.5. Even when fitting the \jone\ flux well (within 10\%), regardless of how many lines are being modeled, the uncertainty in the mass should still be considered to be around 0.4 dex, given variations in the physical conditions and optical depths of the lines. This uncertainty is {\it in addition to} missing the 10\% of mass due to the warm component if mid- to high-J lines are not included. This has implications beyond non-LTE modeling. The conversion from a \jone\ flux to mass using a conversion factor (see Section \ref{sec:alphaco}) should also consider this 0.4 dex uncertainty. In addition to the mass, we also note that without mid- to high-J lines of CO, the total luminosity of the CO (and hence the energetics of the heating/cooling processes) will be significantly underestimated.

The poor fit of a single component to CO \jone\ in over half our galaxies is one indication that the two-component model ought to be used for CO SLED fitting.
To compare the goodness of fit in individual cases, one could compare the reduced chi-squared of the best-fit solution. We encounter a few problems here: first, the best-fit is but one of many well-fitting solutions. More importantly, it is not so straightforward to define the number of degrees of freedom in either the one or two-component model because the parameters (temperature, density, column density, and filling factor) and the resultant model SLED are nonlinear \citep[see][for an excellent discussion]{Andrae2010}. Despite this very imperfect measure, we did attempt to see if we could discover any trends when comparing the reduced chi-squared of the one vs. two component case. For example, are higher-SFR galaxies more likely to require two components for a good fit? We were not able to discern any such trends. 
Some galaxies may simply be able to have their gas described by one component. It is also possible that the uncertainties in the line temperatures, and potential calibration errors comparing low-J to high-J (ground-based to space-based observations, respectively), are other reasons why some galaxies seem to be well modeled by one-component.

While we would generally expect the mass and CO luminosity to scale with \lfir, a more intriguing question is what is the relationship between the excitation conditions of the gas and \lfir\ (SFR)? If bright, highly star-forming galaxies have more gas, is this gas generally excited by the same means and to the same extent as in lower-SFR galaxies? We know within our galaxy that regions of dense, rapid star formation, such as well known molecular associations and the center of the galaxy, have higher gas excitation conditions \citep{Goicoechea2013, Etxaluze2013}.
When observing the total galaxy integrated SLEDs of nearby galaxies, do we see a higher bulk pressure traced by the high-$J$ lines? Or, is this signal ``diluted" by lower-excitation gas, which is more massive and spread over a much larger area of our observing beam?

\begin{figure*}
\includegraphics[width=7in]{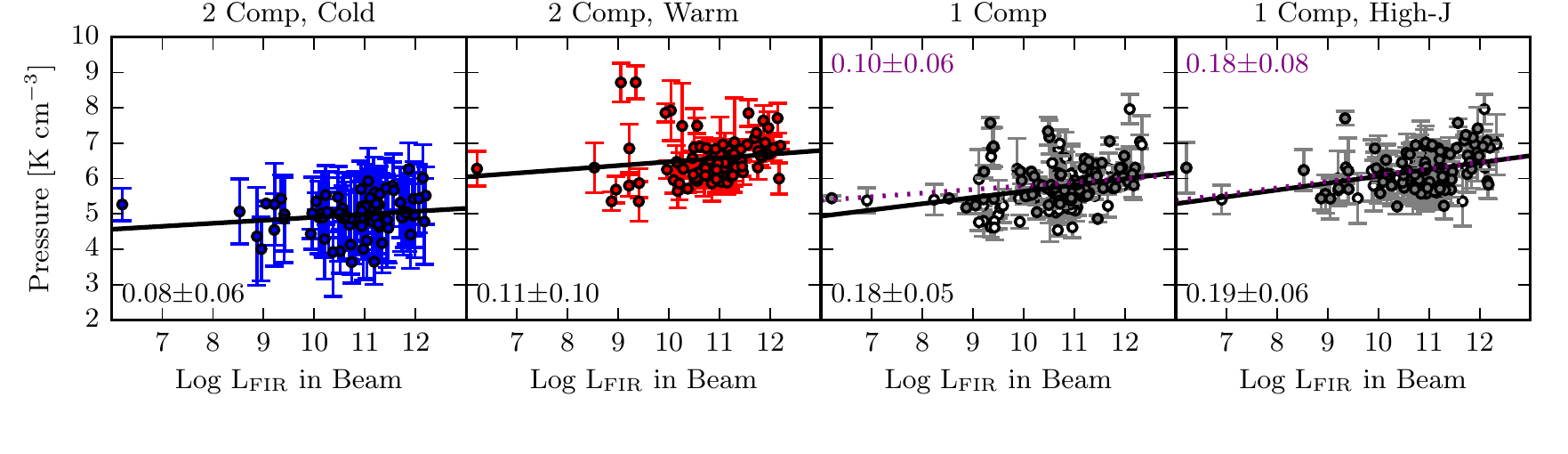} 
\caption{Pressure as a function of \lfir\ in the beam (see Section \ref{sec:obs} for explanation of x-axis). The first two panels from the left are the cold and warm component pressures in the two-component modeling. All these galaxies (87) have at least eight unique CO lines. The next panel is the pressure when modeling the SLEDs with one-component only, for galaxies with at least four unique CO lines (168). The final panel is one-component, modeling only the lines from \jfive\ and up, for galaxies with at least four such lines (128). In the right two panels, filled symbols represent galaxies that also appear in the two component plot. Each panel shows the best-fit line and its slope and error.  In the right two panels, the purple dashed line is a fit to only the filed circles; its slope and error appears at the top of the panel.
\label{fig:pressure}}
\end{figure*}

In Figure \ref{fig:pressure}, we show the pressure in each of our modeling conditions. The first two panels show the two-component model pressures (cold and warm in separate panels). The cold gas is generally around $10^{5.0 \pm 0.5}$ K cm$^{-3}$, with at best a very weak trend with \lfir. 
This indicates that the bulk molecular gas properties, set by the line ratios in the low-J, are fairly uniform across many types of galaxies. 
The warm gas is around $10^{6.5 \pm 0.6}$ K cm$^{-3}$, with little discernible trend as well. 
The lack of a trend in both cold and warm gas pressures means that the bulk of the mass of molecular gas in galaxies in our sample is not affected substantially by the star formation rate. 
The higher star formation rates correlate more explicitly with the quantity of molecular gas than the conditions of that gas. 
We did not find any trend with the ratio of warm to cold mass vs. \lfir.
The implication is that the star formation is not necessarily more efficient in high SFR galaxies than in more quiescent galaxies.
This is further addressed in Section \ref{sec:depletion}.

It is worth noting that the warm and cold emission can ``trade off" to some degree, so long as they add to the same total SLED. This is why the flux likelihoods for the cold and warm components (blue and red shaded areas in left column of Figures \ref{fig:A} through A18) get very wide at high-J for the cold component and low-J for warm component. The warm component emission will be fully responsible for the high-J emission, but the extent to which it is responsible for low-J and high-J can vary depending on how steeply the cold component emission drops off with increasing $J$, which is related to the pressure, and hence this effect is responsible for some of the uncertainty in both pressures.
Only when the SLED shows a noticable hump or discontinuity are the pressures somewhat more fixed; this often causes the high-pressure outliers in the warm component (second from left) panel of Figure \ref{fig:pressure}. 
Some examples include IRASF01417+1651, row 9 in the Appendix figures, and NGC1377, row 16. In these cases, the two-component algorithm has attempted to fit a discontinuity in the SLED with an extremely high-pressure component.
This is why the pressure distributions (middle column) and flux likelihood distributions (left column) are comparatively narrower.
If we instead focus only on the high-J line modeling (rightmost panel in Figure \ref{fig:pressure}, we see those high-pressure outliers go away. 
The aforementioned examples should illustrate why the high-pressure outliers (red in the Appendix figures) are replaced by more moderate pressures in single-component models.
We also find a more noticeable trend, with a slope of $0.19 \pm 0.06$. However, there is still considerable scatter in the warm component pressure. While, in general, the warmest pressures occur in the highest SFR galaxies, this is far from a strong relationship. What this does illustrate, however, is that warm-pressure molecular gas is ubiquitous in nearby galaxies, at least for total \lfir\ $\ge 10^{9.5}$ \ls, though the majority of our galaxies are $\ge 10^{10}$ \ls.

There were many other parameters in our modeling than addressed thus far,
(e.g., temperature, density, area filling factor, ratios of warm to cold luminosity, pressure, or mass), 
but we find no significant relationships between these variables and e.g., \lfir. As mentioned, these parameters such as \tkin\ and \nhtwo are highly degenerate, and so uncertainties on the marginalized parameters are high. 

If there were correlations with \lfir, we would have been able to measure them even given the high uncertainties. We tested this by introducing artificial correlations with \lfir, drawing from the error bar distributions to populate a correlated parameter space, and then fitting the data; we do recover the artificial slopes.
In more detail, we examined log(\lfir) as the independent variable and a given parameter value and error (e.g., log temperature or density), $y \pm \sigma$, as the dependent variable. For slopes $s$ ranging from 0 to 1, we drew a new value, $y_{n} \pm \sigma$ from a gaussian distribution of mean log(\lfir) $\times s$ and width $\sigma$. We then fit $y_{n} \pm \sigma$ as a function of log(\lfir) for 100 iterations of that process. The distributions of the best-fit slopes were centered on the input slope, $s$, with a width of about 0.1 dex. In other words, we should have been able to recover correlations above (or below) zero slope if they were present in the data.
Thus, our conclusions that the temperatures, densities, and ratios of warm/cold pressure, warm/cold luminosity, and warm/cold mass do not correlate with \lfir\ are robust.

\subsection{CO to \mhtwo Conversion Factor}\label{sec:alphaco}

\begin{figure}
\includegraphics[width=\columnwidth]{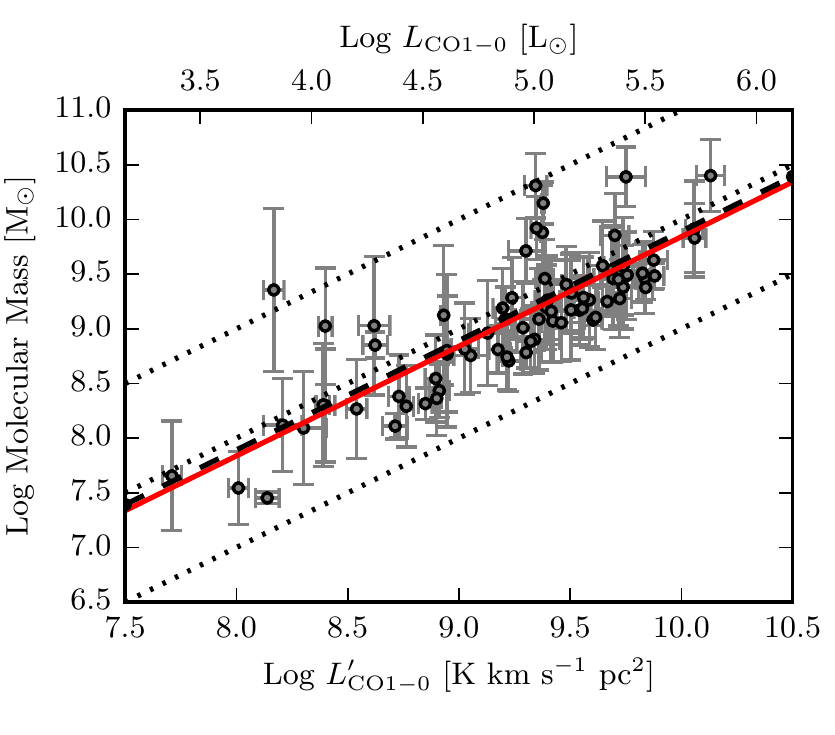} 
\caption{Total molecular mass (warm and cold components) vs. CO \jone\ area-integrated brightness temperature. The axes limits exclude one data point at (X,Y), which does not influence the fits presented. The dotted lines from top to bottom demonstrate \alphaco\ values of 10, 1, and 0.1, respectively. The dashed line is the average \alphaco\ value of 0.8. The solid red line is the best-fit line of slope 1.00 $\pm$ 0.08, which corresponds to a constant \alphaco\ of 0.7.  
\label{fig:alphaco}}
\end{figure}

\begin{figure}
\includegraphics[width=\columnwidth]{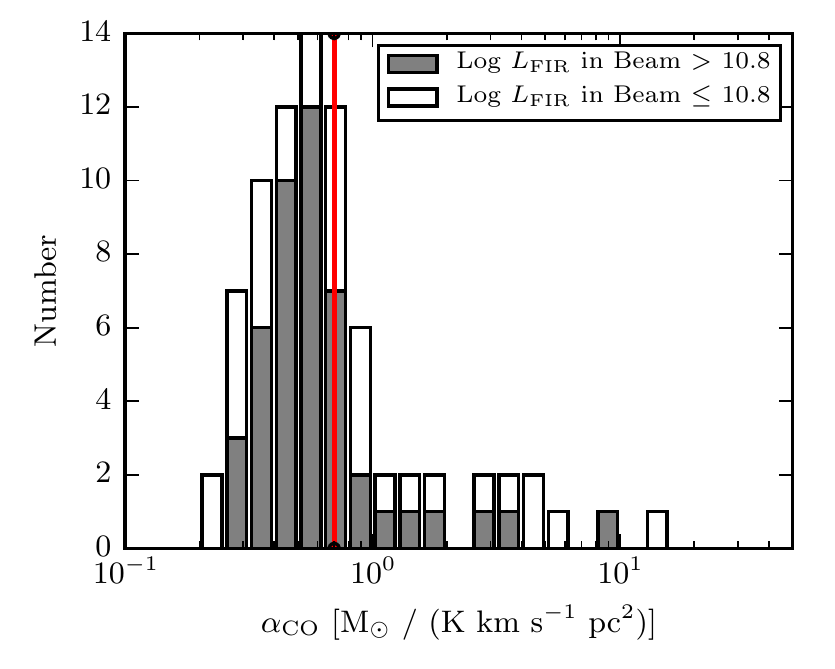} 
\caption{Histogram of \alphaco\ values, the ratio of the y-axis to the x-axis in Figure \ref{fig:alphaco}. The solid red line is the constant \alphaco\ of 0.7 derived from the line fit in that figure. The distribution of \alphaco\ is not centered on this value, but more heavily skewed to slightly lower values. The distribution is not significantly different for those with Log \lfir\ above (gray) or below (white) 10.8 (approximately the cutoff for (U)LIRGs, using 40-120 $\mu$m \lfir\ instead of 8-1000 $\mu$m.
\label{fig:alphaco_hist}}
\end{figure}

Because CO \jone\ is usually in LTE with similar gas conditions from one galaxy to another, the CO \jone\ area-integrated source brightness temperature \lprimeco\ can be directly scaled to total molecular mass \citep{Bolatto2013}. 
The value of the ratio, \alphaco\ = M/\lprimeco, is in units of \ms / (K \kms pc$^2$). 
While the Milky Way has \alphaco $\approx 4$, starburst galaxies and (U)LIRGs may have values 3-10 times lower.

Our estimate of the total molecular mass does not rely on the scaling relation from the \jone\ luminosity: it is calculated using the temperatures and ratios of the first few CO lines (i.e. the shape of the SLED in the first few lines), and takes into account the optical depth of CO in each rotational transition. What we do assume is the abundance of CO to \htwo\ molecules of $10^{-4}$,
and a factor of 1.4 to include He and dust. Our results scale directly with both of these numbers. Therefore, we can plot our derived masses against the CO \jone\ area-integrated brightness temperatures to determine the effective \alphaco\ value for each galaxy. This is shown in Figure \ref{fig:alphaco}. The relationship is well fit by a straight line, which indicates an average \alphaco\ of 0.7 [\ms (K km s$^{-1}$ pc$^{2}$)$^{-1}$]. This value is on the low, but overlapping end of other estimates for similar samples of galaxies. We note that coverage of galaxies below Log(\lfir) = 10$^{10}$ \ls\ is poor, so we are not probing Milky Way-like galaxies. Even our low-\lfir\ galaxies are likely heavily nucleus dominated, which can depress the measured value of \alphaco\ \citep{Bolatto2013}.

As mentioned, changing either of the aforementioned scaling factors can scale this value by a factor of a few.
The majority of the \alphaco\ values are actually slightly lower than this, as can be seen more clearly in Figure \ref{fig:alphaco_hist}, a histogram of the values. There is significant variation in the values derived (standard deviation 0.36 dex), illustrating the uncertainty in applying a single value of \alphaco\ to convert from CO \jone\ luminosity to molecular mass. 

However, this range is consistent with the uncertainty in the cold component mass derived from likelihood modeling. This uncertainty is due to the varying optical depths and excitation temperatures that can simultaneously fit the CO \jone\ flux well (see Figure \ref{fig:results_hist} and associated discussion). We have demonstrated that our method produces masses similar to those using \alphaco\ of 0.7, and quantifies the uncertainty in such masses (using either method) to approximately 0.4 dex.
Methods that rely on scaling the long-wavelength dust continuum to the CO \jone\ line flux should also take this 0.4 dex uncertainty into account when then scaling the dust continuum to another mass measurement \citep{Scoville2016}. 

It is worth noting the importance of the two-component fit in determining this value and greatly reducing the scatter in the relationship. As mentioned above, one-component fits to many CO lines often are fixed by the high-J lines and under-predict the CO \jone\ antenna temperatures and therefore also under-predict the total mass.

\subsection{Gas Depletion Timescales}\label{sec:depletion}

\begin{figure}
\includegraphics[width=\columnwidth]{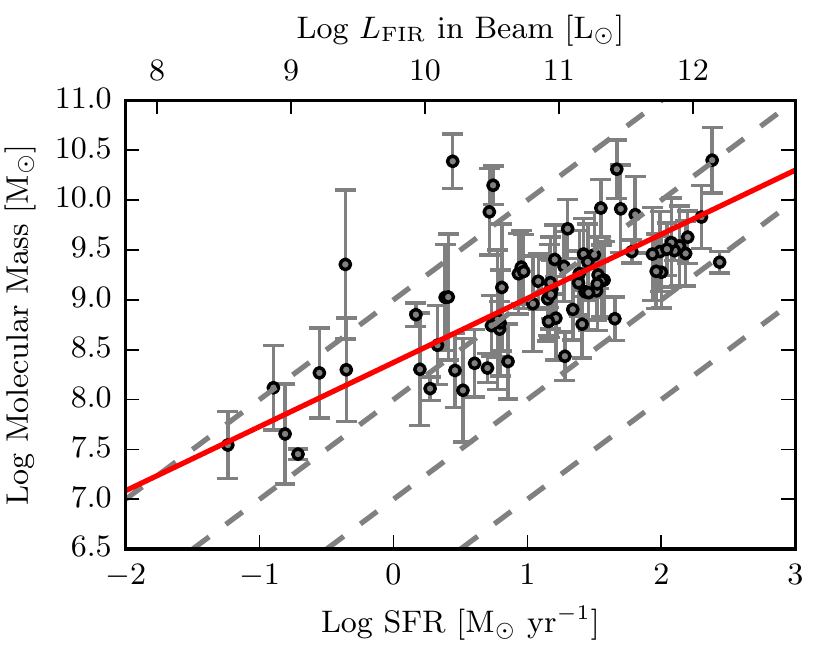} 
\caption{Total molecular mass (warm and cold components) vs. SFR (as directly proportional to \lfir). The axes limits exclude one data point at (X,Y), which does not influence the fits presented. The dotted lines from top to bottom demonstrate \tdep\ values of 1e9, 1e8, 1e7, and 1e6 yr, respectively. The average \tdep value is 1e8 yr. The solid red line is the best-fit line of slope 0.62 $\pm$ 0.09.
\label{fig:tdep}}
\end{figure}

\begin{figure}
\includegraphics[width=\columnwidth]{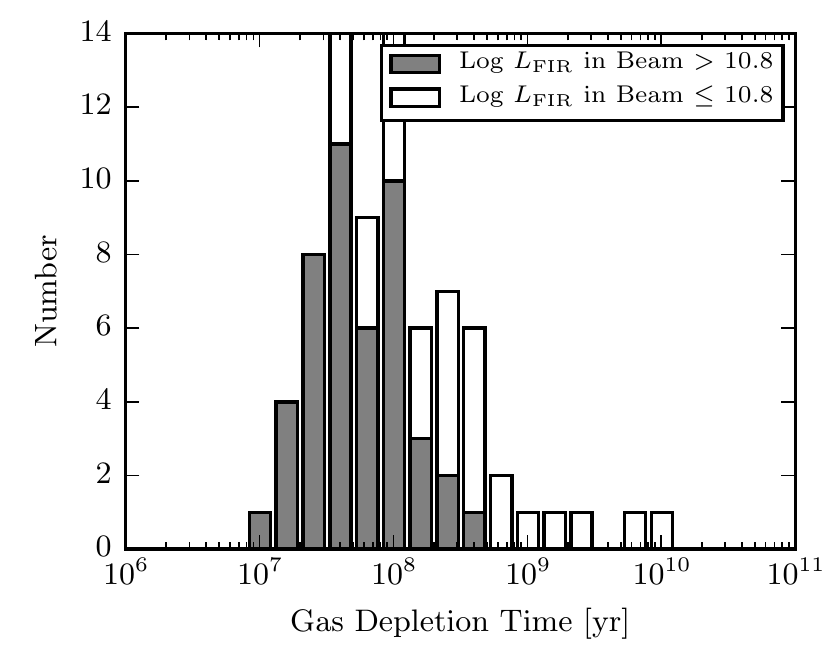} 
\caption{Histogram of \tdep\ values, the ratio of the y-axis to the x-axis in Figure \ref{fig:tdep}.
The distribution is lower for those with Log \lfir\ above (gray) 10.8 (approximately the cutoff for (U)LIRGs, using 40-120 $\mu$m \lfir\ instead of 8-1000 $\mu$m then below 10.8 (white).
\label{fig:tdep_hist}}
\end{figure}

The main sequence (MS) of galaxies refers to the linear variation in SFR with stellar mass (\mstellar), where starburst galaxies lie above this relation \citep[higher SFR per stellar mass][]{Elbaz2011}. 

In other words, galaxies on the MS have a constant specific SFR (sSFR = SFR/\mstellar), which increases with redshift. Stellar masses were available from SDSS for 40 galaxies in our full sample from K14, but only 12 of these had enough CO lines to be modeled here with two components. Furthermore, the estimations for the stellar mass as computed by different methods varied greatly, so we do not show stellar mass (\mstellar), specific SFR (sSFR = SFR/\mstellar), or the molecular gas fraction (\mhtwo / [\mhtwo + \mstellar]). We can, however, compare the molecular gas mass to the SFR, using SFR [\ms\ yr$^{-1}$] = 1.73 $\times 10^{-10}$ \lfir\ [\ls], for the mass and \lfir\ in the beam \citep{Kennicutt1998}, shown in Figure \ref{fig:tdep}. The ratio of these two values is the gas depletion time, \tdep\ [yr], shown also in Figure \ref{fig:tdep_hist}.

We find an average gas depletion time of only $10^{8.0\pm 0.6}$ yr, but a clear trend toward lower gas depletion times in higher-SFR galaxies. In Figure \ref{fig:tdep_hist}, we show the sample separated  into (U)LIRGs (gray) and lower-luminosity galaxies (white). The average depletion times for the two categories are $10^{7.7 \pm 0.4}$ yr and $10^{8.4 \pm 0.6}$ yr, respectively. These results do not support a characteristic, uniform \tdep\ for all these (admittedly varied) galaxies. The (U)LIRGs in our sample have shorter depletion times. We note that this timescale is simply an order of magnitude estimate of what it takes to consume the gas reservoir under a steady SFR, and does not include the effect of SF feedback. The reader should not over-interpret the significance of these depletion timescales, here or in other work.

\citet{Leroy2013} found a fairly constant \tdep\ of 2.2 Gyr with a scatter of 0.3 dex, though with some correlation between \tdep\ and other properties, on 1 kpc scales in nearby disk galaxies. 
This is a factor of 20 longer than our average, but we note some major differences. First, their work focused on disk galaxies, and our sample is heavily weighted towards (U)LIRGs and starburst galaxies. We model with two components only four of their thirty galaxies. (Two of these have \tdep\ $\sim 10^8$, and two $\sim 10^9$ yr$^{-1}$.) Most importantly, they focus on 1 kpc scales, and their average is heavily weighted towards the large outer regions of the disks, which have longer \tdep\ than the inner regions of the galaxies due to inefficient star formation \citep{Bigiel2010}. Our galaxy-integrated data points are likely heavily influenced by the dense central concentrations of star-forming CO gas in galaxies. Treating their sample as galaxy-integrated measurements, they find \tdep\ of 1.3 Gyr, and cite others who find lower values given the same assumptions (e.g. 0.4 Gyr for starbursts).

\citet{Scoville2016} found that high-SFR galaxies at z $\approx 2.2$ and 4.4 have higher gas masses and higher gas mass fractions; in other words, the highly star-forming galaxies are not any more efficient at turning their gas into stars, they simply have more gas (relative to stellar mass) to use. They find gas depletion timescales at z $>$ 1 of 2-7 $\times 10^{8}$ yr, with shorter timescales at higher redshift. 
This value is closer to the efficiencies we find in our largely (U)LIRG population.

It may also be that the dense gas, or more specifically the dense gas fraction, is an important tool for understanding the star formation efficiency in galaxies \citep{Greve2014}. Tracers such as HCN or HCO$^+$ are more sensitive to dense gas than CO. Far infrared color (60 to 100 $\mu$m) may also be an important distinguishing feature for star formation insofar as it provides more information about the dust excitation; \citet{Lu2014} found that C(60/100) is a better predictor of CO SLED shapes than \lfir, and \citet{Leroy2013} found that \tdep\ depends on the dust properties.

\subsection{CO Line Ratios as Diagnostics}

\begin{figure*}
\includegraphics[width=\textwidth]{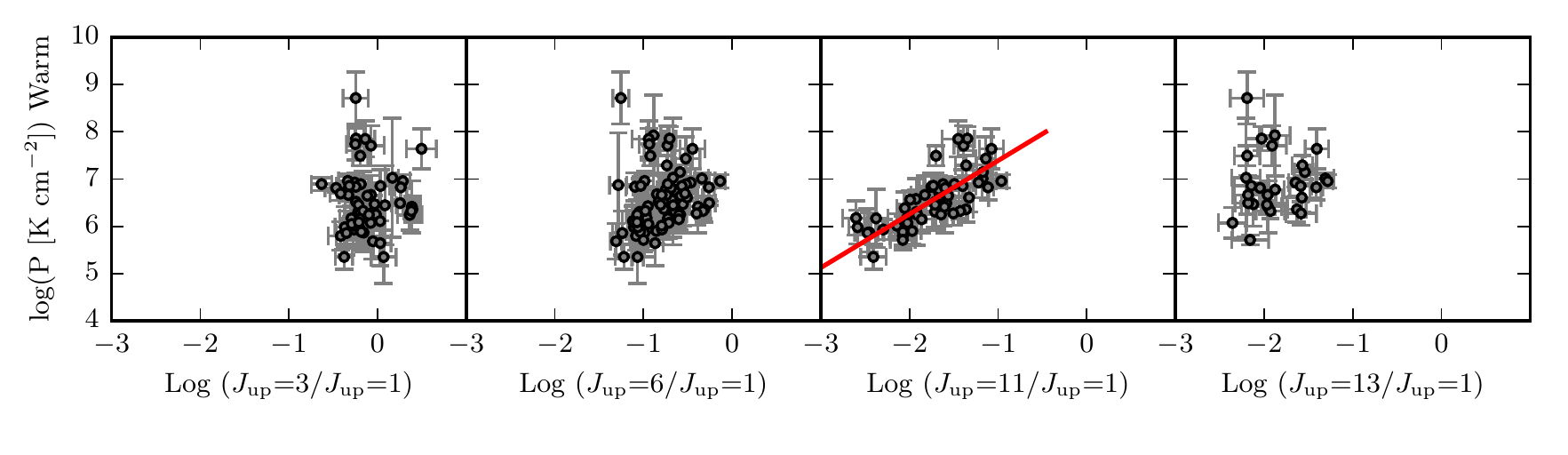} 
\caption{Diagnostic power of selected CO line ratios for warm component gas pressure. The panels contain 57, 78, 54, and 29 data points, from left to right. The third panel from the left includes a best fit of y = (1.13 $\pm 0.15)$ x + (8.52 $\pm 0.27$) from the lnr.bces algorithm by Crist\'obal Sif\'on,which uses the Bivariate, Correlated Errors and intrinsic Scatter method of \citet{Akritas1996}, and takes errors in both dimensions into account. The ratios are formed from the integrated main beam temperatures in units of K km/s.
\label{fig:highjratios}}
\end{figure*}

Without {\it Herschel}, until a similarly wide bandpass instrument is available in space, most studies of the molecular gas conditions in galaxies will not have as many lines available as we use here for modeling. Because of the number of free parameters required (four per component of gas), it is difficult to conduct this same non-LTE excitation modeling without at least eight lines. We sought to determine if there were any line ratios in our survey that 
could serve as diagnostics in lieu of a sufficient number of lines for detailed modeling, 
noting that the high-J CO lines are not in LTE.

The line flux ratios of very close lines span a limited dynamical range across our 87 galaxy sample modeled by two components  (e.g., CO \jsix/\jfive\ has a range of only 0.5 dex). 
In contrast, the dynamical range spanned by the ratio of the highest-J lines to \jone\ is much larger: $\sim$ 1.5 to 2.0 dex.
The range spanned by high-J to mid-J (e.g., \jeleven/\jsix), 1.2 dex, is also large enough to be useful. Figure \ref{fig:highjratios} illustrates the correlation between the warm component pressure and four select line ratios: \jthree, \jsix\, \jeleven, and \jthirteen\ all relative to \jone. As expected, we found that the larger the energy difference between the two lines being compared, the better was its use as a diagnostic of warm component pressure, as we show in the third panel with \jeleven/\jone. We had many fewer data points and greater uncertainty for a higher-$J$ line ratio, such as \jthirteen/\jone, shown in the rightmost panel. 

This is a component of the ISM ripe for study with tools like the Atacama Large Millimeter Array (ALMA).
The highest-J CO line observable for nearby galaxies with ALMA eventually may be \jnine\ in Band 11 
 and for the most part is currently \jsix\ in Band 9 or \jeight\ in Band 10. 
 We find these lines to be well representative of the warm molecular gas, and tracing significantly different molecular gas than e.g., CO \jone.
 As the second panel of Figure \ref{fig:highjratios}, the \jsix\ to \jone\ ratio will not be particularly diagnostic of the warm component pressure. This type of modeling requires more lines to be measured. The \jthirteen\ line and above do lie in the frequency range visible by {\it Herschel}-PACS and now SOFIA; multi-line analysis of Galactic star forming regions may reveal a useful correlation between the warm component pressure and \jthirteen/\jone\ ratio that could be applied to extragalactic sources with limited line measurements. 
Additionally, the line ratios we present here are averages over all the galaxy emission. ALMA observations with high spatial resolution can show variation of the line ratios that we measure here (e.g., \jsix/\jone), especially the differences between the central and outer regions of galaxies.
For higher-redshift galaxies, however, we will be able to see a greater number of CO lines.

\begin{figure*}
\includegraphics[width=\textwidth]{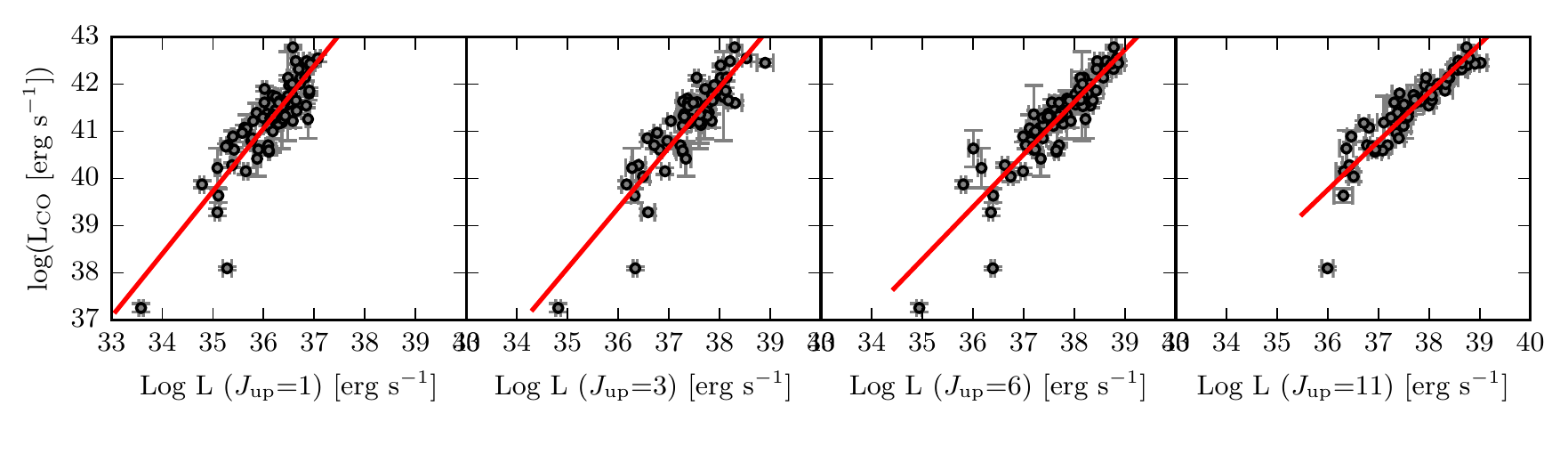} 
\caption{Diagnostic power of selected CO line ratios for total CO luminosity. The panels contain 78, 60, 87, and 62 data points, from left to right.
\label{fig:highjluminosity}}
\end{figure*}

For future reference, we also include comparison of the total CO luminosity (modeled as the warm and cold component sum) vs. four CO individual line luminosities in Figure \ref{fig:highjluminosity}.

\subsection{The Significance of the Detections of a Warm Component of Molecular Gas in Star-Forming, Infrared-Luminous Galaxies}\label{sec:significance}

Prior to the {\it Herschel Space Observatory} there were not a large number of detections of mid and high-$J$ ($J_{upper} \ge 4$) CO lines in galaxies; consequently, most studies of the bulk of the molecular gas using CO as a tracer were focused on cool molecular gas probed by the lowest-lying transitions tens of Kelvin above the ground state.  {\it Herschel's} detection of highly excited CO and a significant (by luminosity and mass) warm component of molecular gas requires an important change in thinking about molecular gas in star-forming galaxies.  Understanding the excitation processes and the consequences for ongoing star formation are going to be a key to understanding feedback into the interstellar medium resulting from star formation. 

The detection of a warm component of molecular gas ($T > 100$ K) in star-forming galaxies by {\it Herschel} is irrefutable and has been clear in a number of works dating from 2010 (see citations in the Section \ref{sec:intro});  And, indeed, recent large-velocity-gradient modeling of {\it Herschel} PACS high-$J$ CO \citep[$J_{upper} \ge 14$, ][]{Mashian2015} 
spectroscopy of many of the same galaxies as in K14 find fits mostly consistent with K14, confirming the presence of a significant component of warm molecular gas.  

However, several caveats about the interpretations of the observations and modeling must be kept in mind.  First, in our sample in particular, and in {\it Herschel} galaxy observations in general, the galaxies with multiple detections of CO transitions of \jfour\ and higher tend to be gas-rich, star-forming galaxies, mostly with infrared luminosities in excess of $10^{10}$ L$_{\odot}$, and the observations are likely dominated by emission from the nuclear regions where the excitation is higher than in more quiescent disk gas.  This leads to a greater sensitivity to and relative importance of warm molecular gas in contrast to, for example, the Milky Way, which by comparison is anemic in highly excited CO.  Second, modeling of CO excitation with RADEX or other large-velocity-gradient approaches is most sensitive to gas pressure and less sensitive to temperature and density independently, which are degenerate and anti correlated in the case of collisional excitation (K14).  Third, and partially caused by the second point, the temperature and density (and mass) uncertainties are typically a factor of several.  Some authors cope with this by stating temperatures and densities that are not unique, or a range of temperatures and densities that are consistent with the observed excitation.  Tying down specific temperatures and densities is difficult and should not be over interpreted.  Our modeling technique generates likelihood distributions that attempt to fully characterize the uncertainties and covariances between parameters, accounting for all known sources of error and parameter degeneracies.  Fourth, {\it Herschel} SPIRE's $\ge 20$ arcsecond diffraction-limited beams undoubtedly integrate ranges of molecular gas physical conditions within galaxies.  

What can be robustly concluded from modeling, in this present study and previous ones, is that in most galaxies at least two molecular gas temperature components are required to explain the low-J, mid-J, and high-J CO line luminosities, with one temperature typically 100 K or greater, often several hundred Kelvin.  Furthermore, the mass of the warm component must be significant:  on order 10\% of the total molecular gas in L$_{IR} > 10^9$ L$_{\odot}$ galaxies.  Indeed, the luminosity-to-mass ratio of the warm molecular gas is $\sim$ 100 times that of the cold molecular gas; if a smaller fraction of the molecular gas were responsible for the warm-component emission, the luminosity-to-mass ratio would have to be even more extreme, causing one to question how CO is not dissociated.  It is also likely the case that there is gas at intermediate temperatures (between the warm and cold components), although model fitting is under constrained with greater than two components, rendering such attempts inconclusive (K14).

It must also be noted that copious warm molecular gas should not come a surprise:  rotational diagrams of far-infrared observations of optically thin H$_2$ have previously demonstrated the presence of warm molecular gas in infrared galaxies \citep[see, e.g.,][]{Lutz2000,Rigopoulou2002,Higdon2006,Armus2007,Roussel2007,Bernard-Salas2009,Brandl2009} 
In fact, in galaxies observed both with H$_2$ and mid and high-J CO in our previous paper (K14), warm molecular gas masses derived from CO and H$_2$ were compatible within the uncertainties (c.f. Tables 12 and 16 of that paper).

An important question that must be addressed observationally and theoretically, is what are the dominant excitation mechanisms of the warm molecular gas?  Conclusions about the excitations mechanisms arise from detailed studies of individual galaxies and vary from galaxy to galaxy, although a picture is beginning to emerge.  Relatively low infrared luminosity star-forming galaxies, such as NGC 4038/4039 \citep{Schirm2014}, have CO spectral line energy distributions consistent with excitation by either photon-dissociation regions (PDRs) or mechanical excitation. The CO spectral line energy distributions of more luminous galaxies, LIRGs and brighter, typically cannot be explained solely by PDRs because the ratio of the brightness of the high-J CO lines (e.g., \jeleven) to lower-J lines (e.g., \jfour)  is too low without extreme densities (i.e., $n > 10^5$ cm$^{-3}$) and the far-infrared emission is too faint given the CO line luminosities.  In some cases, e.g., Arp 220 \citep{Rangwala2011} and M 82 \citep{Panuzzo2010}, cosmic ray excitation is similarly insufficient to excite the warm gas.  Thus, collisional excitation - shocks and turbulent dissipation - seem to be an important contributor to excitation, at least for some of the luminous galaxies, such as NGC6240 \citep{Meijerink2013}.  Further, \citet{Rosenberg2015} suggest that mechanical heating is an important source of excitation of molecular gas in galaxies based on their observed correlation between CO \jone\ line widths and CO excitation in galaxies.  In luminous galaxies with AGN (e.g., NGC 1068), the XDRs are likely partially responsible for the excitation of high-J lines ($J_{upper} > 14$) detected by PACS on {\it Herschel}.   A model consistent with these observations is that the relative importance of shock heating increases with a galaxy's far-infrared luminosity, with the addition that XDRs likely contribute to molecular gas excitation in the case of AGN (see also \citet{Bradford2009}, who find evidence for XDR molecular gas excitation in the case of the Cloverleaf Quasar).

Work prior to {\it Herschel} and {\it Herschel} observations of Galactic molecular gas have also identified warm molecular gas. \citet{Bradford2003} found a warm, dense ($T \sim 120$ K, $n_{H2} \sim 4.5 \times 10^4$ cm$^{-3}$) component of molecular gas in the nucleus of NGC 253 with ground-based CO \jseven\ observations, although the inferred physical conditions were not unique.  They ruled out PDRs as a source of excitation and favored cosmic rays, but did not rule out shocks.  Similarly, \citet{Bradford2005} identified warm, dense molecular gas in the Milky Way's circumnuclear disk and concluded that the gas was heated by magnetohydrodynamic shocks.  With SPIRE PACS and SPIRE spectroscopy from 52 $\mu$m to 671 $\mu$m, \citet{Goicoechea2013} found warm, dense molecular gas in the vicinity of Sgr A* and concluded that the warm gas excitation was likely dominated by shocks or shocks and PDRs, but that PDRs alone were insufficient.  Thus, the case for warm molecular gas, both in star-forming galaxies and extreme regions of the Milky Way, is strong and likely shock excitation plays an important role in heating molecular gas to the warm temperatures observed.

There are at least a few possible mechanisms for shock excitation of the molecular gas:  stellar winds and outflows, supernova remnant and interstellar bubble expansion, and cloud-cloud collisions.  It is currently difficult to distinguish these scenarios, although in the case of Arp 220, the supernova rate has been shown to be high enough to inject sufficient energy into the interstellar medium to explain the warm molecular gas excitation \citep{Rangwala2011}.  However, what remains unclear is how the energy is coupled into the molecular gas.  This will have to be addressed both observationally and with theory and simulations.  
A $J_{\rm upper}$ comfortably greater than 3 is required to probe the warm component. K14 found that \jsix\ luminosity correlated well with far-infrared luminosity and star-formation rate, and it has very little luminosity contribution from cold molecular gas, so it makes a good probe of warm molecular gas (although discriminating PDRs from other excitation mechanisms prefers even higher-J lines).  However, for low-redshift galaxies, \jsix\ is currently the highest-J CO line observable with ALMA and large-scale mapping observations of the type required to study gas excitation as a function of environment will be difficult.  SOFIA has the requisite frequency coverage, but does not currently have the capability for large-scale mapping of high-J CO in the Milky Way or GMCs in nearby galaxies.  On the theoretical front, galaxy and interstellar medium simulations must account for the excitation of the warm molecular gas and assess its consequences for ongoing star formation, such as, e.g., triggering or suppression by dissipation of molecular gas.


\section{Conclusions}\label{sec:conclusions}

The {\it Herschel} SPIRE FTS allowed astronomers to access a largely unseen portion of the electromagnetic spectrum, which specifically traces warm, dense molecular gas. Though the {\it Herschel} FTS could not resolve most extragalactic sources, the sheer number of lines simultaneously detected allowed us to conduct a global census of the molecular gas in many different types of nearby galaxies. 

\begin{itemize}
\item For this sample of galaxies, the low-J CO lines trace molecular gas of pressure $\sim 10^{5.0 \pm 0.5}$ K cm$^{-3}$. The mid- to high-J lines of CO tracer higher-pressure gas at $10^{6.5 \pm 0.6}$ K cm$^{-3}$, and the pressure of this gas is slightly correlated with \lfir. 
\item Two-component fitting is usually but not always necessary to fully fit the \jone\ to \jthirteen\ CO SLEDs of galaxies. 
In about a third of galaxies, modeling the same SLED as one component will not fit the observed CO \jone\ line flux {\it (when including the observed \jone\ line in the fit)}; the \jone\ line flux will be under-predicted by at least a factor of two.
\item If low-J lines are not included in the fit, the mass will be underestimated by an order of magnitude in almost all cases. This is an especially important consideration for the non-LTE modeling of high-redshift galaxies with limited low-J information.
\item If {\it only} low-J lines are fit, the mass is not systematically biased, but the spread when compared to the two-component model of the CO \jone\ through \jthirteen\ SLED is about 0.4 dex.
\item The majority of the luminosity in CO is in the mid- to high-J lines. The median amount missed is 65\%, but some galaxies miss much more. Observing and modeling only the low-J lines will miss this important measure of the energy budget of the gas, which is necessary to understanding the heating mechanisms affecting the CO. This heating cannot be explained by UV PDR models, but is likely mechanical. A more specific characterization cannot ignore mid- to high-J CO lines.
\item Our derived values of \alphaco\ seem linearly constant over this range of \lfir\ at 0.7 [\ms / (K \kms pc$^2$)]. Still, there is considerable spread, with the majority of galaxies (80\%)  having \alphaco\ below 1. 
\item The two aforementioned items indicate that using either method (LTE or \alphaco\ conversion factor), one should consider an uncertainty of about 0.4 dex in a molecular mass estimate from CO. 
\item We find molecular gas depletion timescales (\tdep) of approximately $10^{8}$ yr that decrease with increasing SFR. Our low gas depletion times are likely influenced by the starbursting centers of galaxies.
\item The \jeleven/\jone\ line flux ratio is diagnostic of the warm component pressure. We find log(P [K cm$^{-3}$]) = (1.13 $\pm$ 0.15) $\times$ Log (\jeleven [K \kms]/\jone [K \kms]) +(8.52 $\pm$ 0.27). In general, the greater the difference in energy levels between the two lines, the better one can estimate the warm gas pressure.
\end{itemize}

In conclusion, two-component non-LTE modeling has been conducted for a number of individual targets studied by {\it Herschel}, but this study is unique in the number of galaxies we model, using all available galaxies in the archive. 
We find that the ubiquity (covering $\sim 2.5$ orders of magnitude in \lfir) and luminosity of the warm component of molecular gas were under appreciated in the pre-{\it Herschel} era because of the lack of observational data.
There are few discernible trends in the high-J CO emission with \lfir. In general, higher bulk pressure is found in the higher \lfir\ galaxies.

\section*{Acknowledgements}
This material is based upon work supported by the National Science Foundation under Grant Number AST-1402193 and 
by NASA under award number NNX13AL16G. 
The project was also supported by RSA 1500521 from JPL pursuant to NASA Prime Contract No. NNN12AA01C.
We utilized multiple publicly available software packages in addition to those already credited in the text, such as astropy, astroquery, pyspeckit, and 
Dave Green's ``cubehelix" colormap.
We acknowledge the usage of the HyperLeda database (http://leda.univ-lyon1.fr).
SPIRE has been developed by a consortium of institutes led by Cardiff University (UK) and including Univ. Lethbridge (Canada); 
NAOC (China); CEA, LAM (France); IFSI, Univ. Padua (Italy); IAC (Spain); Stockholm Observatory (Sweden); Imperial College London, 
RAL, UCL-MSSL, UKATC, Univ. Sussex (UK); and Caltech, JPL, NHSC, Univ. Colorado (USA). This development has been supported by national 
funding agencies: CSA (Canada); NAOC (China); CEA, CNES, CNRS (France); ASI (Italy); MCINN (Spain); SNSB (Sweden); STFC (UK); and NASA (USA).



\bibliography{/Users/jkamenetzky/Documents/papers/library}

\appendix

\section{Individual Galaxy SLEDs and Parameter Distributions}\label{sec:indv}

In this appendix, we present the individual results for all of the SLEDs modeled as two components. Table \ref{tbl:results} presents the means and uncertainties in the marginalized parameter distributions (for the most probable mode in likelihood space, if multiple modes were found) for the cold and warm pressures, masses, and luminosities. 

\begin{figure*}
    \centering
    \includegraphics[width=\textwidth]{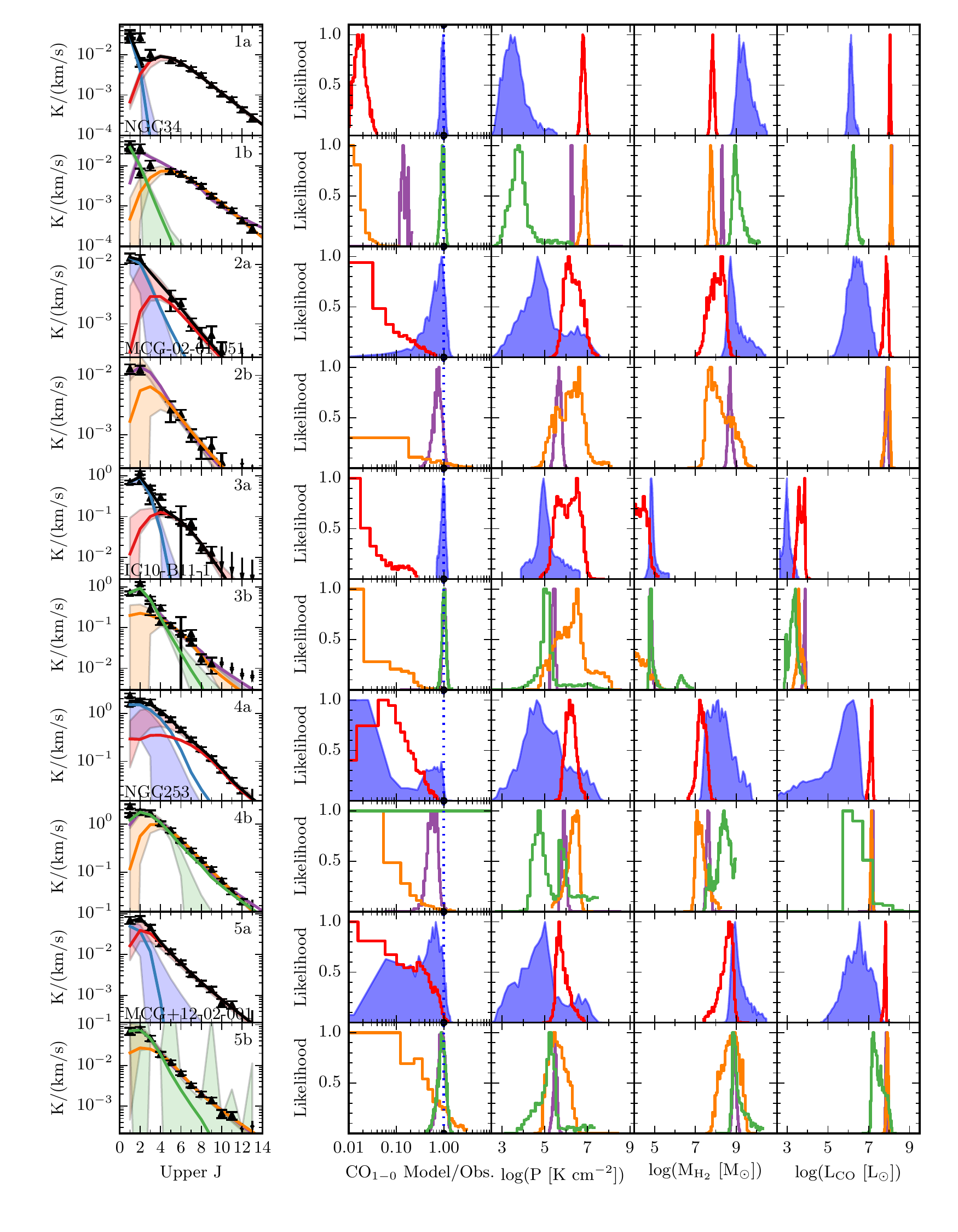}
    \caption{Individual Galaxy Results. See Appendix text for explanation.
    \label{fig:A}}
\end{figure*}

In Figures \ref{fig:A} through A18, we present the modeled SLEDs and the likelihood distributions for the aforementioned parameters.  
Figures A2 to A18 are available online.
The galaxies in this set of figures are listed in order by RA and Dec, in the same order as Table 1 of \citet{Kamenetzky2016} (skipping ones without at least 8 unique CO lines). 
Each galaxy occupies two rows: the top row ("a") shows the 2-component model (cold=blue, warm=red), and the bottom row ("b") shows the three different 1-component models: one component (all lines, purple), one component high-J only (orange), and if available, one component low-J only (green). 
Each row is numbered for reference in the text.
 The leftmost panel shows the modeled SLED (black points), including the best fit (solid line) and 1$\sigma$ range of modeled fluxes (shaded region).
 The bottom left of the panel shows the galaxy name, and the top right shows the row in this set of figures, for reference in the text. 
 The right four panels show the marginalized likelihoods for the CO \jone\ flux relative to the weighted mean of the measured fluxes (if \jone\ available, otherwise this panel is blank), the effective pressure, the total molecular mass, and the total CO luminosity, respectively. The colors are the same as in the SLEDs, though the cold component is also shaded.
The warm \jone\ flux often does not appear in the 2nd column if it is too faint. All histograms have 40 bins, but their size depends on the size of the parameter distribution, so some appear more finely sampled than others.

The first galaxy in our list, NGC34 (top of Figure \ref{fig:A}) nicely illustrates the different modeling paradigms. Row 1a shows the modeling results for the 2-component case, where blue represents the cold component, and red represents the warm component. 
If one modeled the same lines using only one component (purple in row 1b), the result would underestimate the CO \jone\ flux (first and second columns), be of slightly lower but similar pressure than the 2-component warm pressure (third column), be of lower mass than the 2-component cold mass (fourth column), and be of similar luminosity to the 2-component warm component (fifth column). Modeling only the low-J lines (green in row 1b) would produce similar results to the cold component in the 2-component modeling, but underestimate the mass slightly. Finally, if one only modeled the high-J lines (orange in row 1b), one would severely underestimate the CO \jone\ flux and mass (first, second, fourth columns), and find similar pressure and luminosity to the 2-component warm component (third, fifth columns).

Some SLEDs have some unusual features. As previously mentioned, some show a ``hump" which separates the two components very distinctly (IRASF01417+1651, row 9, and NGC1377, row 16).
IRAS09022-3615 (row 30) does not have any low-J CO lines available, so there is little data with which to fit the cold component (other than the requirement that the cold component be more massive than the warm component). 

Most of our galaxies are well-fit by one mode, or region in the multi-dimension parameter space. Of our two-component fits, 11/87 (13\%) are instead fit well by two modes in parameter space. This means there are two separate, distinct areas of parameter space with measurable likelihood. Some examples include NGC3256 (row 35) and IC4687 (row 72). Two distinct peaks in some marginalized parameters are visible. Multiple ($>$1) modes in parameter space are more likely in the one-component modeling cases: 57\% of 1-component (full SLED), 27\% of 1-comp high-J only, and 42\% of 1-comp low-J only.

Sometimes the best-fit solution may not be representative of the likelihood space, which demonstrates the importance of examining the marginalized parameter distributions instead of simply relying on one best-fit solution when non-LTE modeling. Some examples include NGC253 (Row 4a) and NGC1482 (row 17a).

When a SLED is extremely poorly fit (especially by one component), the best-fit and/or the likelihood distribution of the SLED fluxes (shaded regions in first column of appendix figures) may be influenced by parameter space subject to computational problems in the matrix solving of level populations and optical depth.
Discontinuities like seen in green (low-J only) in row 5b (MGC+12-02-001) are a result of this; the best fit (solid line) is physically meaningful, but the marginalized distributions of the fluxes (shaded regions) were clearly subject to errors. Slight discontinuities can also be found in IRAS01417+1641 single-comp, full SLED (row 9b, purple). In the case of NGC1377 (Row 16b), the best-fit SLEDs have negative excitation temperatures and optical depths for the e.g. \jthree\ and \jfour\ lines. 
Such lines are not used in the calculation of the likelihood.

\begin{landscape}
\begin{table}
\footnotesize
\caption{Two Component Likelihood Results}
\label{tbl:results}
\begin{tabular}{lrrrrrrrrrrrrrrr}
\hline
\# & Galaxy & RA & Dec & \multicolumn{2}{c}{$P_{\rm Cold}$} & \multicolumn{2}{c}{$P_{\rm Warm}$} & \multicolumn{2}{c}{$M_{\rm H_2, Cold}$} & \multicolumn{2}{c}{$M_{\rm H_2, Warm}$}  & \multicolumn{2}{c}{$L_{\rm CO, Cold}$} & \multicolumn{2}{c}{$L_{\rm CO, Warm}$} \\
 & &  &  & \multicolumn{2}{c}{[K cm$^{-3}$]} & \multicolumn{2}{c}{[K cm$^{-3}$]} & \multicolumn{2}{c}{[\ms]} & \multicolumn{2}{c}{[\ms]}  & \multicolumn{2}{c}{[erg s$^{-1}$]} & \multicolumn{2}{c}{[erg s$^{-1}$]} \\
 & & J2000 & J2000 & Mean & $\sigma$ & Mean & $\sigma$ & Mean & $\sigma$  & Mean & $\sigma$ & Mean & $\sigma$  & Mean & $\sigma$\\
\hline
1 & NGC34 & 0h11m06.55s & $-$12d06m26.3s & 3.6 & 0.6 & 6.8 & 0.1 & 9.4 & 0.4 & 7.8 & 0.1 & 39.73 & 0.13 & 41.63 & 0.03 \\
2 & MCG-02-01-051 & 0h18m50.86s & $-$10d22m37.5s & 4.9 & 1.0 & 6.4 & 0.4 & 9.0 & 0.5 & 8.1 & 0.4 & 40.01 & 0.44 & 41.43 & 0.10 \\
3 & IC10-B11-1 & 0h20m27.70s & +59d16m59.4s & 5.3 & 0.5 & 6.3 & 0.5 & 4.9 & 0.3 & 4.0 & 0.4 & 36.56 & 0.11 & 37.16 & 0.09 \\
4 & NGC253 & 0h47m33.12s & $-$25d17m17.6s & 5.0 & 1.0 & 6.3 & 0.2 & 8.3 & 0.6 & 7.3 & 0.2 & 38.99 & 1.00 & 40.70 & 0.08 \\
5 & MCG+12-02-001 & 0h54m03.61s & +73d05m11.8s & 4.8 & 1.0 & 5.9 & 0.3 & 9.2 & 0.4 & 8.5 & 0.4 & 40.00 & 0.59 & 41.38 & 0.06 \\
6 & NGC0317B & 0h57m40.37s & +43d47m32.4s & 5.7 & 0.8 & 6.8 & 0.3 & 8.8 & 0.1 & 7.4 & 0.2 & 40.38 & 0.14 & 41.22 & 0.11 \\
7 & IC1623 & 1h07m47.00s & $-$17d30m25.0s & 5.6 & 0.3 & 7.0 & 1.3 & 9.2 & 0.1 & 7.0 & 1.2 & 41.66 & 0.28 & 40.97 & 0.94 \\
8 & CGCG436-030 & 1h20m02.58s & +14d21m42.5s & 4.8 & 1.2 & 6.3 & 0.3 & 9.1 & 0.5 & 8.4 & 0.3 & 40.06 & 0.63 & 41.75 & 0.11 \\
9 & IRASF01417+1651 & 1h44m30.52s & +17d06m08.9s & 5.4 & 0.8 & 7.7 & 0.3 & 8.9 & 0.3 & 7.4 & 0.1 & 40.48 & 0.15 & 41.57 & 0.07 \\
10 & NGC1068 & 2h42m40.71s & $-$00d00m47.8s & 5.2 & 0.4 & 7.5 & 0.2 & 8.7 & 0.3 & 6.8 & 0.2 & 40.84 & 0.15 & 41.23 & 0.09 \\
11 & UGC02369 & 2h54m01.81s & +14d58m14.3s & 4.7 & 1.1 & 6.1 & 0.5 & 9.4 & 0.4 & 8.5 & 0.5 & 40.24 & 0.57 & 41.64 & 0.08 \\
12 & NGC1266 & 3h16m00.70s & $-$02d25m38.0s & 4.4 & 0.1 & 7.9 & 0.3 & 8.8 & 0.1 & 6.6 & 0.1 & 40.17 & 0.06 & 40.89 & 0.02 \\
13 & 3C 84 & 3h19m48.16s & +41d30m42.1s & 3.6 & 0.6 & 6.9 & 0.2 & 9.3 & 0.4 & 7.4 & 0.2 & 39.57 & 0.13 & 41.31 & 0.06 \\
14 & NGC1365-SW & 3h33m35.90s & $-$36d08m35.0s & 5.0 & 1.3 & 6.1 & 0.3 & 9.9 & 0.4 & 7.9 & 0.3 & 39.02 & 0.71 & 41.13 & 0.05 \\
15 & NGC1365-NE & 3h33m36.60s & $-$36d08m20.0s & 3.9 & 0.6 & 6.0 & 0.4 & 10.1 & 0.2 & 8.1 & 0.3 & 40.13 & 0.20 & 41.13 & 0.05 \\
16 & NGC1377 & 3h36m39.10s & $-$20d54m08.0s & 5.4 & 0.7 & 8.7 & 0.5 & 7.2 & 0.6 & 5.2 & 0.1 & 39.61 & 0.12 & 40.59 & 0.39 \\
17 & NGC1482 & 3h54m38.90s & $-$20d30m09.0s & 3.9 & 1.3 & 5.7 & 0.1 & 8.2 & 0.3 & 7.8 & 0.1 & 37.89 & 0.47 & 40.62 & 0.03 \\
18 & NGC1614 & 4h33m59.85s & $-$08d34m44.0s & 4.9 & 1.3 & 6.2 & 0.3 & 9.9 & 0.3 & 8.4 & 0.3 & 39.50 & 0.83 & 41.60 & 0.05 \\
19 & IRAS F05189-2524 & 5h21m01.47s & $-$25d21m45.4s & 5.1 & 1.0 & 6.8 & 0.3 & 9.2 & 0.4 & 8.2 & 0.3 & 40.41 & 0.48 & 42.00 & 0.05 \\
20 & MCG+08-11-002 & 5h40m43.65s & +49d41m41.8s & 4.9 & 1.0 & 6.1 & 0.3 & 9.3 & 0.4 & 8.6 & 0.3 & 40.26 & 0.48 & 41.60 & 0.08 \\
21 & NGC1961 & 5h42m04.37s & +69d22m41.9s & 5.1 & 0.6 & 7.5 & 1.2 & 9.1 & 0.3 & 7.0 & 0.8 & 40.90 & 0.27 & 41.17 & 0.61 \\
22 & IRAS 06035-7102 & 6h02m54.01s & $-$71d03m10.2s & 5.4 & 1.1 & 7.4 & 0.5 & 9.6 & 0.3 & 8.5 & 0.2 & 41.09 & 0.64 & 42.41 & 0.07 \\
23 & NGC2146-NW & 6h18m36.70s & +78d21m32.0s & 4.9 & 0.9 & 5.9 & 0.2 & 8.7 & 0.5 & 7.9 & 0.2 & 39.34 & 0.85 & 40.54 & 0.07 \\
24 & NGC2146-nuc & 6h18m38.60s & +78d21m24.0s & 4.9 & 1.2 & 6.2 & 0.3 & 9.1 & 0.6 & 7.8 & 0.2 & 38.42 & 0.92 & 40.70 & 0.04 \\
25 & NGC2146-SE & 6h18m40.50s & +78d21m16.0s & 5.0 & 1.0 & 6.0 & 0.2 & 8.8 & 0.7 & 7.8 & 0.2 & 39.15 & 0.94 & 40.58 & 0.08 \\
26 & NGC2369 & 7h16m37.60s & $-$62d20m35.9s & 4.1 & 0.8 & 6.1 & 0.4 & 9.3 & 0.4 & 8.1 & 0.3 & 39.97 & 0.20 & 41.27 & 0.05 \\
27 & NGC2388a & 7h28m53.43s & +33d49m08.4s & 4.7 & 0.9 & 6.0 & 0.4 & 9.1 & 0.5 & 8.2 & 0.4 & 39.97 & 0.45 & 41.24 & 0.10 \\
28 & MCG+02-20-003 & 7h35m43.44s & +11d42m34.8s & 4.9 & 1.3 & 6.5 & 0.6 & 8.9 & 0.5 & 7.6 & 0.5 & 39.77 & 0.53 & 41.10 & 0.13 \\
29 & NGC2623 & 8h38m24.08s & +25d45m16.6s & 4.2 & 0.8 & 6.7 & 0.1 & 9.2 & 0.4 & 8.0 & 0.1 & 39.84 & 0.22 & 41.69 & 0.03 \\
30 & IRAS09022-3615 & 9h04m12.72s & $-$36d27m01.3s & 4.8 & 1.2 & 6.0 & 0.4 & 11.1 & 1.2 & 9.5 & 0.5 & 41.19 & 0.77 & 42.42 & 0.05 \\
31 & NGC2798 & 9h17m22.90s & +41d59m59.0s & 5.0 & 1.2 & 6.3 & 0.3 & 8.0 & 0.5 & 7.4 & 0.3 & 39.00 & 0.72 & 40.67 & 0.06 \\
32 & UGC05101 & 9h35m51.65s & +61d21m11.3s & 4.9 & 1.1 & 6.3 & 0.3 & 9.4 & 0.4 & 8.7 & 0.3 & 40.41 & 0.52 & 41.97 & 0.07 \\
33 & M82 & 9h55m52.22s & +69d40m46.9s & 5.0 & 1.2 & 6.5 & 0.5 & 9.0 & 0.5 & 7.2 & 0.4 & 38.72 & 0.75 & 40.61 & 0.04 \\
34 & NGC3227 & 10h23m30.58s & +19d51m54.2s & 4.5 & 1.0 & 5.8 & 0.3 & 8.2 & 0.5 & 7.4 & 0.3 & 38.80 & 0.42 & 40.27 & 0.05 \\
35 & NGC3256 & 10h27m51.27s & $-$43d54m13.8s & 5.6 & 0.6 & 6.6 & 0.2 & 9.0 & 0.0 & 8.1 & 0.2 & 41.05 & 0.16 & 41.68 & 0.05 \\
36 & NGC3351 & 10h43m57.70s & +11d42m14.0s & 5.3 & 0.1 & 8.7 & 0.6 & 7.5 & 0.1 & 4.8 & 0.2 & 39.64 & 0.04 & 40.09 & 0.42 \\
37 & IRASF10565+2448 & 10h59m18.17s & +24d32m34.4s & 5.0 & 0.9 & 6.6 & 0.3 & 9.5 & 0.3 & 8.6 & 0.2 & 40.73 & 0.38 & 42.12 & 0.07 \\
38 & NGC3627 & 11h20m15.00s & +12d59m30.0s & 5.3 & 1.2 & 6.9 & 0.7 & 7.8 & 0.7 & 6.4 & 0.5 & 38.77 & 0.77 & 40.02 & 0.10 \\
39 & Arp299-B & 11h28m31.00s & +58d33m41.0s & 5.3 & 1.1 & 6.4 & 0.3 & 9.0 & 0.3 & 8.3 & 0.3 & 40.39 & 0.69 & 41.71 & 0.05 \\
40 & Arp299-C & 11h28m31.00s & +58d33m50.0s & 4.8 & 1.2 & 6.3 & 0.2 & 9.1 & 0.4 & 8.4 & 0.2 & 39.99 & 0.68 & 41.65 & 0.04 \\
41 & Arp299-A & 11h28m33.63s & +58d33m47.0s & 5.7 & 0.8 & 6.8 & 0.3 & 8.7 & 0.2 & 8.0 & 0.3 & 40.64 & 0.45 & 41.87 & 0.05 \\
42 & ESO 320-G030 & 11h53m11.72s & $-$39d07m48.9s & 4.9 & 1.1 & 6.2 & 0.3 & 8.7 & 0.4 & 8.1 & 0.3 & 39.79 & 0.66 & 41.28 & 0.07 \\
43 & NGC4051 & 12h03m09.61s & +44d31m52.8s & 5.1 & 0.9 & 6.3 & 0.7 & 7.5 & 0.3 & 6.5 & 0.7 & 38.61 & 0.59 & 39.59 & 0.14 \\
44 & NGC4194 & 12h14m09.63s & +54d31m36.1s & 4.9 & 1.0 & 6.9 & 0.1 & 8.4 & 0.4 & 7.0 & 0.1 & 39.25 & 0.19 & 40.84 & 0.08 \\
\hline
\end{tabular}
\end{table}

\clearpage

\begin{table}
\footnotesize
\contcaption{Two Component Likelihood Results}
\label{tbl:results}
\begin{tabular}{lrrrrrrrrrrrrrrr}
\hline
\# & Galaxy & RA & Dec & \multicolumn{2}{c}{$P_{\rm Cold}$} & \multicolumn{2}{c}{$P_{\rm Warm}$} & \multicolumn{2}{c}{$M_{\rm H_2, Cold}$} & \multicolumn{2}{c}{$M_{\rm H_2, Warm}$}  & \multicolumn{2}{c}{$L_{\rm CO, Cold}$} & \multicolumn{2}{c}{$L_{\rm CO, Warm}$} \\
 & &  &  & \multicolumn{2}{c}{[K cm$^{-3}$]} & \multicolumn{2}{c}{[K cm$^{-3}$]} & \multicolumn{2}{c}{[\ms]} & \multicolumn{2}{c}{[\ms]}  & \multicolumn{2}{c}{[erg s$^{-1}$]} & \multicolumn{2}{c}{[erg s$^{-1}$]} \\
&  & J2000 & J2000 & Mean & $\sigma$ & Mean & $\sigma$ & Mean & $\sigma$  & Mean & $\sigma$ & Mean & $\sigma$  & Mean & $\sigma$\\
\hline
45 & IRAS12116-5615 & 12h14m22.17s & $-$56d32m32.8s & 4.9 & 1.2 & 6.3 & 0.6 & 10.0 & 1.3 & 8.4 & 0.6 & 40.51 & 0.83 & 41.64 & 0.31 \\
46 & NGC4388 & 12h25m46.75s & +12d39m43.5s & 4.9 & 1.0 & 6.0 & 0.4 & 8.5 & 0.4 & 7.8 & 0.4 & 39.37 & 0.59 & 40.78 & 0.09 \\
47 & NGC4536 & 12h34m27.00s & +02d11m17.0s & 5.3 & 0.2 & 7.9 & 0.8 & 8.1 & 0.1 & 6.2 & 0.5 & 40.40 & 0.18 & 40.96 & 0.29 \\
48 & Mrk 231 & 12h56m14.23s & +56d52m25.2s & 5.5 & 0.8 & 6.9 & 0.1 & 9.3 & 0.1 & 8.5 & 0.1 & 40.78 & 0.19 & 42.39 & 0.04 \\
49 & NGC4826 & 12h56m43.70s & +21d40m58.0s & 4.9 & 1.2 & 5.4 & 0.6 & 9.4 & 0.7 & 7.4 & 0.5 & 38.09 & 0.67 & 39.26 & 0.08 \\
50 & ESO507-G070 & 13h02m52.34s & $-$23d55m17.8s & 5.3 & 1.0 & 6.7 & 0.4 & 9.0 & 0.4 & 7.9 & 0.4 & 40.43 & 0.47 & 41.58 & 0.07 \\
51 & IRAS13120-5453 & 13h15m06.35s & $-$55d09m22.7s & 5.0 & 1.2 & 6.7 & 0.2 & 10.3 & 1.2 & 8.7 & 0.2 & 40.74 & 0.82 & 42.31 & 0.09 \\
52 & Arp193 & 13h20m35.34s & +34d08m22.2s & 4.8 & 1.3 & 6.5 & 0.0 & 10.3 & 0.3 & 8.4 & 0.0 & 38.50 & 0.12 & 41.76 & 0.02 \\
53 & Cen A & 13h25m27.61s & $-$43d01m08.8s & 4.0 & 0.9 & 5.7 & 0.4 & 7.6 & 0.5 & 6.5 & 0.3 & 38.63 & 0.31 & 39.85 & 0.07 \\
54 & NGC5135 & 13h25m44.06s & $-$29d50m01.2s & 4.8 & 0.9 & 5.9 & 0.5 & 9.1 & 0.3 & 8.3 & 0.7 & 40.27 & 0.53 & 41.21 & 0.41 \\
55 & ESO 173-G015 & 13h27m23.78s & $-$57d29m22.2s & 5.0 & 1.2 & 6.7 & 0.2 & 9.3 & 1.0 & 8.0 & 0.2 & 39.82 & 0.77 & 41.78 & 0.04 \\
56 & M83 & 13h37m00.92s & $-$29d51m56.7s & 5.0 & 1.0 & 5.9 & 0.4 & 8.3 & 0.5 & 7.2 & 0.4 & 38.96 & 0.68 & 40.13 & 0.07 \\
57 & Mrk 273 & 13h44m42.11s & +55d53m12.7s & 4.4 & 0.9 & 7.0 & 0.1 & 9.5 & 0.4 & 8.5 & 0.1 & 40.15 & 0.23 & 42.49 & 0.06 \\
58 & IRAS 14348-1447 & 14h37m38.26s & $-$15d00m24.6s & 5.4 & 1.0 & 6.9 & 0.3 & 9.8 & 0.3 & 8.9 & 0.3 & 41.16 & 0.47 & 42.53 & 0.07 \\
59 & NGC5713 & 14h40m11.50s & $-$00d17m20.0s & 4.9 & 1.1 & 5.6 & 0.5 & 9.0 & 0.6 & 7.8 & 0.7 & 38.90 & 0.94 & 40.41 & 0.37 \\
60 & IRAS 14378-3651 & 14h40m59.01s & $-$37d04m32.0s & 5.0 & 1.1 & 6.6 & 0.3 & 9.5 & 0.4 & 8.8 & 0.3 & 40.51 & 0.60 & 42.30 & 0.08 \\
61 & CGCG049-057 & 15h13m13.09s & +07d13m31.8s & 5.3 & 0.8 & 6.9 & 0.2 & 8.4 & 0.2 & 7.5 & 0.1 & 40.02 & 0.21 & 41.19 & 0.04 \\
62 & VV705 & 15h18m06.13s & +42d44m44.5s & 4.2 & 1.0 & 6.5 & 0.2 & 9.5 & 0.4 & 8.4 & 0.2 & 40.05 & 0.29 & 41.86 & 0.07 \\
63 & ESO099-G004 & 15h24m57.99s & $-$63d07m30.2s & 5.2 & 1.2 & 6.7 & 1.0 & 9.6 & 1.3 & 8.0 & 0.9 & 40.76 & 0.80 & 41.59 & 0.21 \\
64 & Arp220 & 15h34m57.12s & +23d30m11.5s & 5.0 & 0.9 & 6.7 & 0.1 & 9.4 & 0.3 & 8.4 & 0.1 & 40.50 & 0.24 & 42.13 & 0.05 \\
65 & CGCG052-037 & 16h30m56.60s & +04d04m58.3s & 5.5 & 0.8 & 6.4 & 0.8 & 8.9 & 0.3 & 7.8 & 0.9 & 40.66 & 0.75 & 41.25 & 0.51 \\
66 & NGC6156 & 16h34m52.50s & $-$60d37m07.7s & 5.0 & 0.7 & 6.9 & 1.1 & 8.7 & 0.3 & 7.1 & 0.9 & 40.35 & 0.51 & 40.89 & 0.44 \\
67 & IRASF16399-0937 & 16h42m40.10s & $-$09d43m13.6s & 4.3 & 1.1 & 5.9 & 0.5 & 10.4 & 0.3 & 8.5 & 0.5 & 40.17 & 0.32 & 41.41 & 0.05 \\
68 & NGC6240 & 16h52m58.89s & +02d24m03.4s & 5.8 & 0.7 & 7.0 & 0.1 & 9.4 & 0.1 & 8.8 & 0.1 & 41.52 & 0.29 & 42.76 & 0.04 \\
69 & IRASF17138-1017 & 17h16m35.82s & $-$10d20m41.5s & 5.1 & 0.8 & 6.2 & 0.6 & 9.0 & 0.4 & 8.0 & 0.6 & 40.41 & 0.57 & 41.27 & 0.19 \\
70 & IRAS F17207-0014 & 17h23m21.96s & $-$00d17m00.9s & 6.0 & 0.9 & 7.7 & 0.4 & 10.4 & 0.3 & 8.4 & 0.1 & 41.44 & 0.16 & 42.45 & 0.05 \\
71 & IRAS17578-0400 & 18h00m31.86s & $-$04d00m53.3s & 5.9 & 0.9 & 7.0 & 0.1 & 9.7 & 0.3 & 7.3 & 0.1 & 40.06 & 0.12 & 41.30 & 0.08 \\
72 & IC4687 & 18h13m39.63s & $-$57d43m31.3s & 4.8 & 0.9 & 6.7 & 0.1 & 9.0 & 0.4 & 7.6 & 0.1 & 40.08 & 0.22 & 41.13 & 0.04 \\
73 & IRAS F18293-3413 & 18h32m41.13s & $-$34d11m27.5s & 4.6 & 1.0 & 6.2 & 0.4 & 9.9 & 0.4 & 8.7 & 0.3 & 40.56 & 0.22 & 41.78 & 0.04 \\
74 & IC4734 & 18h38m25.60s & $-$57d29m25.1s & 4.0 & 0.9 & 6.4 & 0.2 & 9.4 & 0.3 & 7.9 & 0.2 & 39.87 & 0.12 & 41.37 & 0.07 \\
75 & NGC6701 & 18h43m12.56s & +60d39m11.3s & 4.7 & 1.0 & 6.3 & 0.8 & 9.2 & 0.4 & 7.9 & 0.8 & 40.12 & 0.63 & 41.19 & 0.14 \\
76 & NGC6946 & 20h34m52.30s & +60d09m14.0s & 4.4 & 1.4 & 5.4 & 0.3 & 8.0 & 0.4 & 7.3 & 0.3 & 36.71 & 0.47 & 38.08 & 0.04 \\
77 & CGCG448-020 & 20h57m24.33s & +17d07m38.3s & 5.7 & 1.0 & 7.8 & 0.4 & 9.9 & 0.4 & 7.7 & 0.1 & 40.95 & 0.20 & 42.10 & 0.13 \\
78 & ESO286-IG019 & 20h58m26.79s & $-$42d39m00.6s & 5.3 & 1.0 & 7.1 & 0.2 & 9.4 & 0.5 & 8.2 & 0.1 & 40.82 & 0.55 & 42.31 & 0.05 \\
79 & NGC7130 & 21h48m19.50s & $-$34d57m04.7s & 4.2 & 1.1 & 5.9 & 0.3 & 9.3 & 0.4 & 8.6 & 0.3 & 39.66 & 0.50 & 41.54 & 0.04 \\
80 & IRAS 22491-1808 & 22h51m49.26s & $-$17d52m23.5s & 6.3 & 0.7 & 7.6 & 0.4 & 9.5 & 0.1 & 8.3 & 0.2 & 41.70 & 0.28 & 42.37 & 0.08 \\
81 & NGC7469 & 23h03m15.62s & +08d52m26.4s & 5.1 & 1.0 & 6.1 & 0.3 & 9.1 & 0.3 & 8.5 & 0.4 & 40.37 & 0.79 & 41.54 & 0.16 \\
82 & ESO148-IG002 & 23h15m46.72s & $-$59d03m15.1s & 4.9 & 1.0 & 7.3 & 0.2 & 9.3 & 0.4 & 7.9 & 0.1 & 40.34 & 0.38 & 42.00 & 0.06 \\
83 & IC5298 & 23h16m00.64s & +25d33m23.7s & 4.6 & 0.9 & 6.7 & 0.1 & 9.1 & 0.5 & 8.0 & 0.1 & 39.93 & 0.21 & 41.52 & 0.07 \\
84 & NGC7552 & 23h16m10.77s & $-$42d35m05.4s & 5.5 & 0.7 & 6.6 & 0.2 & 8.3 & 0.1 & 7.4 & 0.2 & 40.25 & 0.24 & 41.01 & 0.05 \\
85 & NGC7582 & 23h18m23.50s & $-$42d22m14.0s & 5.5 & 0.8 & 6.4 & 0.3 & 8.2 & 0.4 & 7.4 & 0.4 & 40.01 & 0.58 & 40.83 & 0.12 \\
86 & NGC7771 & 23h51m24.88s & +20d06m42.6s & 5.2 & 0.8 & 6.2 & 0.6 & 9.1 & 0.3 & 7.9 & 0.9 & 40.63 & 0.57 & 41.05 & 0.56 \\
87 & Mrk331 & 23h51m26.80s & +20d35m09.9s & 5.4 & 1.2 & 6.4 & 0.5 & 8.7 & 0.3 & 8.0 & 0.6 & 40.27 & 0.97 & 41.36 & 0.20 \\
\hline
\end{tabular}
\end{table}

\end{landscape}
\clearpage

\centering

\bsp	
\label{lastpage}
\end{document}